\documentclass{article}
\usepackage{fullpage}
\usepackage{amsmath}
\usepackage{amsthm}
\usepackage{amsfonts}
\usepackage{booktabs}
\usepackage{hyperref}
\usepackage{algorithm}
\usepackage{algpseudocode}
\usepackage{graphicx}
\usepackage{subcaption}
\usepackage{multirow}
\usepackage{arydshln}
\usepackage{amssymb}
\usepackage{tikz}
\usepackage{float}
\usepackage[dvipsnames,table]{xcolor}
\graphicspath{{./images/}}
\begin{document}
		
	\title{Disentangling Popularity and Quality: An Edge Classification Approach for Fair Recommendation}

	\author{Nemat Gholinejad, Mostafa Haghir Chehreghani\\
	Department of Computer Engineering\\
	Amirkabir University of Technology (Tehran Polytechnic)\\
	Tehran, Iran\\
	\texttt{\{n.gholinezhad,mostafa.chehreghani\}@aut.ac.ir}}
	\date{}
	\maketitle
	
	\begin{abstract}
		Graph neural networks (GNNs) have proven to be an effective tool for enhancing the performance of recommender systems. However, these systems often suffer from popularity bias, leading to an unfair advantage for frequently interacted items, while overlooking high-quality but less popular items.
		In this paper, we propose a GNN-based recommendation model that disentangles popularity and quality to address this issue. Unlike existing methods that treat all long-tail items uniformly, our approach introduces an edge classification technique to differentiate between popularity bias and genuine quality disparities among items. Furthermore, it uses cost-sensitive learning to adjust the misclassification penalties, ensuring that underrepresented yet relevant items are not unfairly disregarded.
		Experimental results demonstrate improvements in fairness metrics by approximately $32\%$ on average across different scenarios while maintaining competitive accuracy, with only minor variations compared to state-of-the-art methods.
	\end{abstract}
	
	\vspace{\baselineskip}
	
	\section{Introduction}
\label{sec:introduction}

Recommender systems have become a cornerstone of the digital experience,
guiding users through vast amounts of content by suggesting items such as movies, books, or products based on their preferences.
As these systems have evolved,
Graph Neural Networks (GNNs)~\cite{gcmc,10.1145/3700790,DBLP:journals/tjs/ZohrabiSC24,DBLP:journals/natmi/Chehreghani22,lightgcn,pinsage} have emerged as a powerful approach for capturing complex user-item interactions, due to their ability to model complex and nonlinear structures and relationships within graph-based data. However, recent research in GNN-based recommendation continues to rely on the same well-known Bayesian Personalized Ranking (BPR)~\cite{bpr} objective function previously used in traditional methods~\cite{lightgcn, ngcf, dgrec, dgcf}. BPR operates on the principle that observed interactions should be ranked higher than unobserved ones. However, this method tends to reinforce the popularity of already well-known items, further marginalizing long-tail items and effectively treating all un-interacted items as irrelevant.
Users might not be aware of the existence of many of these items. If they were to encounter them, they might prefer some over those they have already seen.
This problem leads to a situation where items with high previous visit rates are increasingly promoted by the recommender system, while less-visited items remain overlooked.
This issue is commonly referred to as exposure bias or popularity bias in the literature.
The item degree distribution in recommender systems follows a power law distribution, where a small number of items have a large number of interactions, while the majority have only a few interactions. The former group is referred to as short-head or popular items, while the latter is known as long-tail or unpopular items.

Moreover, despite the existence of popularity bias or exposure bias, not all items in the non-interacted category suffer from this issue. Some items simply lack the quality needed to be recommended as desirable by recommender systems~\cite{zhao2022popularity}. Current recommender systems designed to address this bias~\cite{adjnorm, apda, popularitymetrics, hetrofair} treat all long-tail items equally. Objective functions such as BPR also do not have a way to distinguish between these two categories.

To address these challenges, in this paper, we propose a novel approach that directly tackles the exposure bias introduced by the BPR objective function through an edge classification framework. This method reclassifies the edges within the user-item interaction graph, distinguishing between high-quality and low-quality long-tail items. Our approach ensures that popularity bias is mitigated without sacrificing the recommendation of genuinely relevant items. Furthermore, we enhance this framework with cost-sensitive learning, which adjusts the misclassification penalties, particularly for unpopular items. This adjustment encourages the model to fairly recommend items that might otherwise be unjustly overlooked due to their lack of popularity.

We distinguish between popularity caused by item quality and popularity caused by the model's tendency to interact with popular items. This distinction enhances user-centricity by helping users get closer to their true preferences. By introducing edge classification and cost-sensitive learning together, we ensure that less popular items in a user’s profile are not wrongly considered negative. Our fairness improvement stems not from recommending less popular items to users with mostly popular items, but from recommending more relevant items to those with a profile made up of less popular items. The final evaluation, based on the equal chance method, focuses on users’ preferences and interests to assess the model’s performance in improving fairness, demonstrating its effectiveness.

We conduct experiments on several well-known datasets and demonstrate that our proposed model significantly outperforms state-of-the-art methods in terms of fairness metrics, achieving improvements in fairness measures by approximately 32\% on average compared to existing approaches. Moreover, it achieves competitive accuracy, with only minor variations compared to the best existing methods, occasionally outperforming them and in other cases showing a slight decline. This indicates that our method balances the trade-off between fairness and accuracy,
effectively. Additionally, through ablation studies, we highlight the importance of the fairness-oriented component of our model.

The choice of an edge classification framework is driven by the limitations of the commonly used BPR objective, which treats all unobserved interactions as negative and thus implicitly reinforces popularity bias. By reformulating recommendation as a binary edge classification problem, we remove this pairwise ranking assumption and allow the model to evaluate each interaction independently. This provides a more flexible representation, particularly for long-tail items that may be relevant but underexposed.
Building on this, the cost-sensitive learning component is introduced to account for asymmetric misclassification risks, specifically reducing the likelihood of incorrectly suppressing high-quality long-tail items. Together, these design choices enable the model to address the underlying source of popularity bias while maintaining control over the fairness–accuracy trade-off through the hyperparameter $\lambda$.

The structure of this paper is as follows: Section~\ref{sec:relatedwork} offers a brief overview of related work, followed by the necessary preliminaries and definitions in Section~\ref{sec:preliminaries}. Section~\ref{sec:ourmethod} presents a detailed description of our proposed method. In Section~\ref{sec:experiments}, we discuss the results of our extensive experiments, highlighting the performance of the proposed method. Section~\ref{sec:limitations} discusses the limitations of our approach and outlines future work. Finally, Section~\ref{sec:conclusion} concludes the paper.

\section{Related work}
\label{sec:relatedwork}

In this section, we explore recent developments in three areas closely connected to our work: GNN-based recommendation, data-driven fair recommendation, and strategies for mitigating popularity bias.

\subsection{GNN-based recommendation}

The success of GNN-based methods in graph-related tasks has led to their application in recommender systems. GCMC~\cite{gcmc} uses this technique for matrix completion tasks on user-item bipartite graphs. NGCF~\cite{ngcf} is the first attempt to acquire users' and items' representations based on multi-hop neighborhoods. GTN~\cite{gtn} adaptively captures the reliability of user-item interactions, aiming to enhance the robustness of recommendations by mitigating the impact of unreliable behaviors. Wei and Chow~\cite{wei2023fgcr} propose the Fused Graph Context-aware Recommender system (FGCR), which integrates GCN and traditional models to enhance user-item-context interactions. They introduce a masked graph convolution strategy that improves information aggregation across different types of nodes.
Peng et al.~\cite{peng2024powerful} address limitations in GCN-based recommendation systems by introducing generalized graph normalization to handle noise across different data densities and an individualized graph filter to enhance expressive power. After introducing SGC~\cite{wu2019simplifying}, which shows that non-linearity is an unnecessary function in GCN~\cite{gcn}, LRGCCF~\cite{lrgccf} and LightGCN~\cite{lightgcn} both work on simplifying NGCF inspired by SGC. The former removes non-linearity from NGCF, while the latter removes feature transformation in addition to non-linearity. The only trainable parameters of LightGCN are the nodes' initial embeddings.
Recently, Peng et al.~\cite{peng2024less} identified three key redundancies in GNN-based recommendation methods: feature, structure, and distribution redundancies. The authors propose a Simplified Graph Denoising Encoder that reduces complexity by using only the top-$K$ singular vectors and introduces a scalable contrastive learning framework to enhance model robustness.

\subsection{Data-driven fair recommendation}
This approach tries to improve fairness by modifying the training data.
Research in this domain is limited. Ekstrand et al.~\cite{ekstrand2018all} use demographic features to divide users into different groups and apply re-sampling to modify the distribution of various user groups within the training dataset. Rastegarpanah et al.~\cite{rastegarpanah2019fighting} tackle user-side unfairness by adding antidote data to the training data, introducing unreal user nodes. The fairness objective function is optimized using gradient descent to update the augmented antidote data. Chen et al.~\cite{chen2023improving} introduce a model-agnostic framework that improves fairness through data augmentation, generating synthetic user-item interactions based on the hypothesis that users with different sensitive attributes may have similar item preferences. However, most existing works have focused on user-side fairness, while research on item-side fairness using a data-driven approach is very rare.

\subsection{Popularity debiasing}
Fair recommendation is closely related to popularity bias, which refers to the tendency of recommender systems to favor popular items, often at the expense of long-tail items. In recent years, various methods have been proposed to tackle this problem. SGL~\cite{sgl} introduces a self-supervised setting for GNN-based recommendation, creating multiple views for each node in the graph and using contrastive learning to maximize agreement between these views. Zhu et al.~\cite{popularitymetrics} propose  a regularization term to achieve a balanced recommendation list between popular and long-tail items, introducing two metrics to evaluate bias from both the user and item sides. The r-AdjNorm method~\cite{adjnorm} adjusts the power of the symmetric square root normalization term in graph neural networks to control the normalization process during neighborhood aggregation, aiming for improved results, especially for low-degree items. APDA~\cite{apda} assigns lower weights to connected edges during the aggregation process and employs residual connections to achieve unbiased and fair representations for users and items in graph collaborative filtering. Anelli et al.~\cite{anelli2023auditing} analyze both consumer and producer fairness in the graph collaborative filtering approach. Zhao et al.~\cite{zhao2022popularity} differentiate popularity bias into benign and harmful categories, arguing that item quality can influence popularity bias and introducing a time-aware disentangled framework to identify and mitigate harmful bias. Chen et al.~\cite{chen2024graph} theoretically assess how graph convolution exacerbates popularity bias, demonstrating that stacking multiple graph convolution layers accelerates users' gravitation towards popular items in the representation space. 

\section{Preliminaries}
\label{sec:preliminaries}
In this section, we introduce the key notations and definitions used throughout the paper, followed by an overview of the Light Graph Convolutional (LightGCN) model~\cite{lightgcn}. Since our method leverages LightGCN to generate users' and items' embeddings as input for the edge classification module, we briefly review its formulation here.

\subsection{Notations and definitions}

In graph collaborative filtering based recommender systems, historical interactions between users and items are formulated as an undirected bipartite graph $G = (V, E)$,
where $V = U \cup I$. Here, \(U = \{u_1,u_2,\ldots,u_{|U|}\}\) is a set of users and \(I=\{i_1,i_2,\ldots, i_{|I|}\}\) is a set of items.
The term bipartite means that there are no internal edges within the user group or the item group. Instead, users have interactions with different items.
These interactions consider all user behaviors, such as buying, clicking, viewing, etc.
These interactions are represented by $E$.
The adjacency matrix $A$ of the graph $G$ is defined as $A \in \mathbb{R}^{\left(\left|U\right| + \left|I\right|\right) \times \left(\left|U\right| + \left|I\right|\right)}$,
where $A_{ui} = A_{iu} = 1$ if $(u,i) \in E$ and otherwise $0$.

\subsection{Long-tail distribution}
\begin{figure}[!t]
	\centering
	\includegraphics[width=0.5\linewidth]{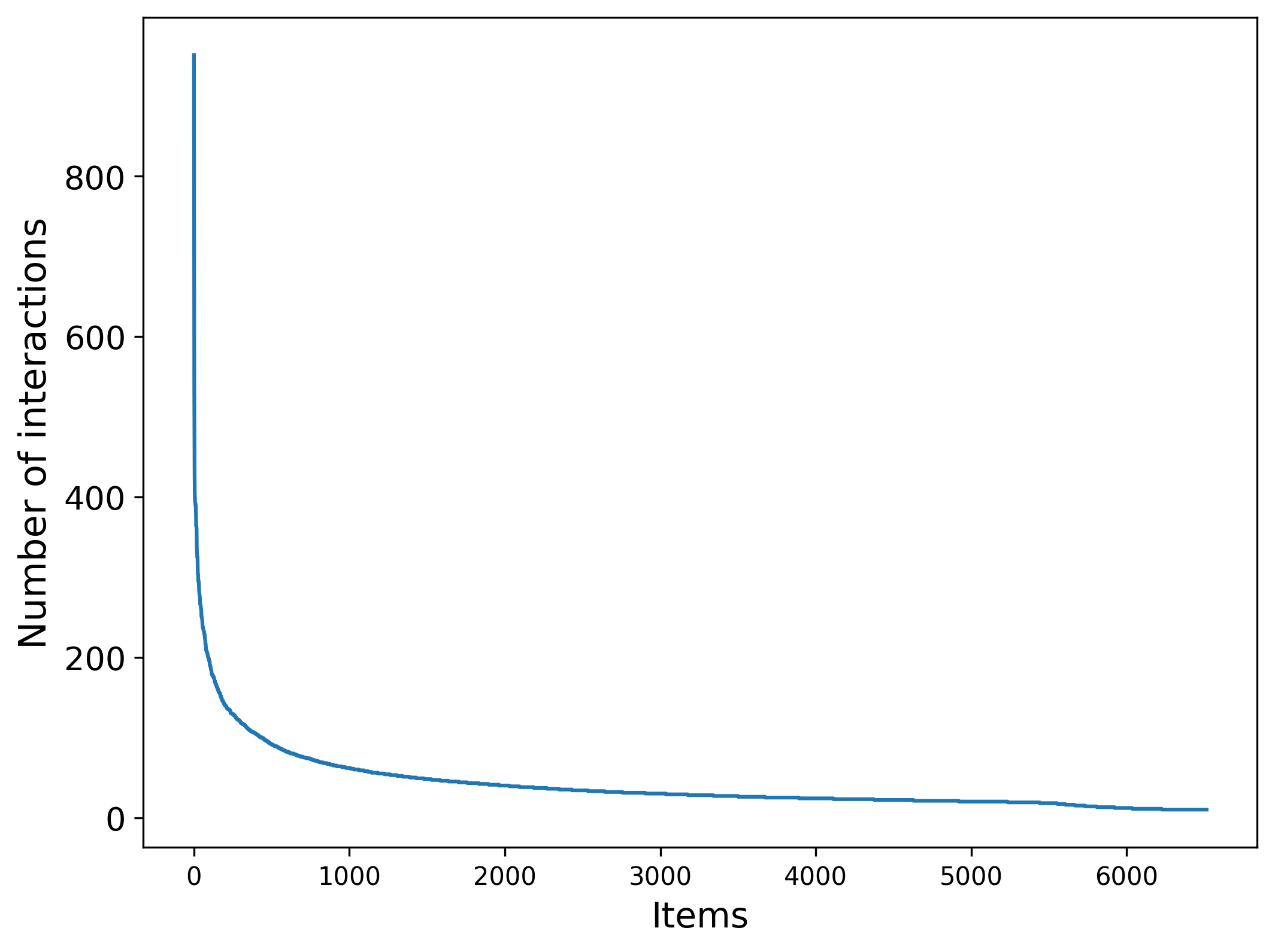}
	\caption{Degree distribution of items in the Bookcrossing dataset.\label{fig:powerlaw}}
\end{figure}

A power law degree distribution is a pattern where the majority of items have very few occurrences, while a small number of items have a large number of occurrences. This type of distribution is common in many systems, especially where certain items are much more popular than others. In the context of recommender systems, the item degree distribution follows this  power law pattern. A small group of items, often called short-head or popular items, receive the majority of user interactions, meaning they are frequently recommended and interacted with. On the other hand, the majority of items, referred to as long-tail or unpopular items, receive far fewer interactions and tend to be less visible to users. This imbalance in item interactions is a typical challenge in many recommendation systems, where a few items dominate user attention, while most items remain relatively undiscovered. This phenomenon can be visually represented through a long-tail distribution curve, where the short-head items create a steep initial drop, followed by a long, gradually declining tail that represents the many less popular items. Figure~\ref{fig:powerlaw} demonstrates this curve, highlighting the stark contrast between the highly popular items and the vast number of less popular items.

\subsection{Light graph convolutional model}

The light graph convolution used in our method is directly derived from LightGCN \cite{lightgcn}. It is one of the pioneering studies in graph collaborative filtering simplifies GCN  due to the nature of collaborative filtering graphs, where nodes do not have rich initial features like those in citation networks such as Cora or CiteSeer. The simplification process involves removing two components: non-linearity and feature transformation. Therefore, the standard GCN changes to the following form~\cite{lightgcn}: 
\begin{equation}
	H^{\left(k+1\right)} = (D^{-\frac{1}{2}}AD^{-\frac{1}{2}})H^{\left(k\right)},
	\label{eq:lightgcn_prop}
\end{equation}
where \(D\)  is the diagonal degree matrix that has the same dimension as
the adjacency matrix \(A\) of the graph, \(D_{ii} = \sum_{j} A_{ij}\), and
$H^{\left(k\right)}$ is the embedding matrix at layer $k$. 
\(H^{\left(0\right)} = [h_{1}^{\left(0\right)}, h_{2}^{\left(0\right)}, \ldots, h_{|U|+|I|}^{\left(0\right)} ]^T\) is the initial embedding matrix of
the nodes in the collaborative filtering graph, and $h^{k} \in \mathbb{R}^{d\times1}.$

\section{Our proposed method}
\label{sec:ourmethod}

In this section, we first introduce our method for disentangling popularity and quality in GNN-based recommendation systems. We then describe our edge classification technique, followed by our approach to cost-sensitive learning.

\subsection{Disentangled popularity and quality}
\label{disentangled}

Not all items that fall into the long tail part of the item degree distribution are wrongly discriminated against by popularity bias. Some of these items simply do not have the necessary quality to be recommended by recommender systems. Therefore, we must categorize long-tail items in terms of quality and distinguish between popularity bias that causes high-quality items to be ignored and bias that filters out low-quality items. For this purpose, we use a combination of the baseline estimation~\cite{riccif2011handbook} method and the degree of nodes (here, items) to identify low-quality items.
The baseline estimation method is as follows:
\begin{equation}
	b_{ui} = \mu + b_{u} + b_{i},
	\label{eq:baseline}
\end{equation}
where $b_{ui}$ is the baseline estimation for user $u$'s rating on item $i$,
$\mu$ is the average rating over all items, and
$b_{u}$ and $b_{i}$ indicate the observed deviations of user $u$ and item $i$, respectively and are defined as:
\begin{align}
	b_{u} &= \bar{r}_{u} - \mu, \label{eq:b_u}\\
	b_{i} &= \bar{r}_{i} - \mu, \label{eq:b_i}
\end{align}
where $\bar{r}_{u}$ is the average rating given by user $u$ and $\bar{r}_{i}$ is the average rating received by item $i$~\cite{riccif2011handbook}.
For each edge $(u, i) \in E$, we calculate the baseline estimation and determine the difference between this value and the score given by user $u$ to item $i$, which we call the error. Specifically, we subtract the baseline estimation from the actual score. A positive number indicates that user $u$ is interested in item $i$, as the score is higher than the baseline estimation based on $\mu$, $b_u$, and $b_i$. Conversely, a negative number indicates a relative lack of interest. After calculating the baseline estimation, we remove the edge $(u, i)$ from the adjacency matrix if less than two-thirds of users have given a positive score to item $i$ and if item $i$ has a degree of less than $\gamma$.
In general, we set $\gamma$ to $20$, but we study the effect of this hyperparameter in Section~\ref{sec:gamma_effect}.

\noindent It is crucial to emphasize that not all popularity is inherently negative. Distinguishing between quality and popularity allows us to remove noise from the graph and optimally combat that part of popularity which is caused by popularity bias. This approach results in a more fair model.  Importantly, we avoid treating all low-degree and poor-quality items as victims of popularity bias. Simply recommending these items without considering their actual quality would lower user satisfaction and worsen the user experience, as users may be presented with irrelevant or low-quality content. By balancing this, we ensure that recommendations are fair without compromising the quality and relevance of the items recommended to the user.

The threshold of two-thirds positive scores is chosen to balance precision and recall in identifying low-quality items. Empirically, this threshold effectively filters items that consistently receive negative feedback while preserving those with generally positive reception. However, we acknowledge that this rule may inadvertently remove edges associated with items that appeal to a small but distinct user subgroup, which raises a potential user-side fairness concern. In such cases, niche-interest items may be under-represented in the filtered graph and consequently under-recommended to their intended audience.
To assess this effect, we additionally evaluate a less restrictive threshold (0.5), and report the results in the experimental section. The findings indicate that while a lower threshold retains more niche items, it also introduces more noise from low-quality interactions, highlighting a trade-off between preserving minority-interest items and filtering unreliable signals.
In practice, this risk can be mitigated by (1) setting $\gamma$ to focus filtering on extremely low-degree items, and (2) leveraging the cost-sensitive component to upweight remaining long-tail items. Exploring adaptive or user-aware thresholding strategies is an important direction for future work.

\subsection{Edge classification}
\label{sec:edgeclassification}

Although implicit feedback-based classification is a special kind of edge classification where all existing edges are labeled as $1$, we can extend it to binary classification with the help of negative sampling. Therefore, we can provide the following definition:

\newtheorem{definition}{Definition}
\begin{definition}
	\label{thm:edgeclassificationdefiniton}
	In an undirected bipartite graph $G=\left(V,E\right)$ with users $U$ and items $I$ and known positive interactions $(u,i) \in E$ the edge classification problem in recommender systems involves predicting whether unknown edges $(u,i) \notin E$, should be classified as $1$ indicating potential positive interactions or $0$ indicating no interaction.
\end{definition}

\noindent The reason for using this method instead of the well-known BPR method is investigated in recent research~\cite{nipsunbiased, damak2021debiased, apda}. BPR, a widely used pairwise objective function for recommendation, has been shown to be vulnerable to popularity or exposure bias in multiple studies. It assumes that any positive interaction should be ranked higher than all other unobserved potential interactions.
Satio~\cite{nipsunbiased} demonstrates that the estimator optimized in the BPR algorithm is biased against the ideal pairwise loss. Zhou et al.~\cite{apda} investigate how the gradient magnitude of the BPR algorithm causes the interest scores of users and items to be positively correlated with the degree of the aggregated nodes in the $k$-hop neighborhood.
To tackle this problem, we introduce a new edge classification setting for the recommender system, as defined in Definition~\ref{thm:edgeclassificationdefiniton}, and use cross-entropy as the objective function, instead of BPR. The binary cross-entropy can be defined as follows:
\begin{equation}
	L = - \sum_{i \in I} y_{ui} \log \tilde{y}_{ui} + \left(1- y_{ui}\right) \log\left(1- \tilde{y}_{ui}\right).
	\label{eq:bce}
\end{equation}
Here, $y_{ui}$ can be considered a label: it will be $1$ if user $u$ has item $i$ in his profile and $0$ otherwise. $\tilde{y}_{ui}$ is the model-predicted score (interest score) and is in the range of $\left[0,1\right]$. This loss function is also known as the negative log-likelihood loss.
Not suppressing items that could potentially attract the user's attention increases the likelihood of those items being noticed, which can improve user satisfaction. Based on the degree distribution of items, most of the items in the negative category are less popular. Therefore, the limitations imposed by the BPR method cause the user to miss many items they might like, simply due to a lack of past interactions. To address this, we use an edge classification approach to improve the user experience by increasing the diversity and relevance of recommendations, giving less popular but relevant items a chance to be discovered. In other words, we do not filter items solely based on the lack of past visits, aiming to build user trust in the process.

While the proposed quality filtering step relies on explicit ratings (e.g., to compute $\bar{r}_i$ and $b_i$), the core components of our framework, edge classification and cost-sensitive learning, are not restricted to explicit-feedback settings. For purely implicit data (e.g., clicks, views, or purchases), the baseline estimator can be replaced with an implicit preference proxy, such as interaction frequency, dwell-time signals, or confidence-weighted scores derived from user engagement. In this case, the notion of a "positive score" (e.g., two-thirds of users providing positive feedback) can be reinterpreted using thresholds on implicit signals, such as the proportion of users exceeding a minimum engagement level. Therefore, while the current formulation assumes explicit ratings, the overall framework remains applicable, provided that an appropriate quality proxy is defined for the available feedback signals. We further discuss this limitation and possible adaptations in Section~\ref{sec:limitations}.

\subsection{Cost-sensitive learning}
\label{sec:constsensitive}

Misclassification costs are not the same in supervised machine-learning tasks; different misclassification errors incur different penalties~\cite{elkan2001foundations, ijcnncost}. For example, the cost of failing to diagnose an incurable disease is much higher than misclassifying a negative case as positive. This issue also occurs in recommender systems. In systems affected by popularity bias, unpopular items are often erroneously considered false negatives. Despite their relevance, these items have fewer interactions, making them susceptible to popularity bias and thus not recommended to users.  In our recommendation model, we address this issue as follows:
Let class one denote items that the user interacted with in the past (positive interactions), and class zero refers to negative (un-interacted) items.
We proportionally increase the misclassification cost for class zero and decrease it for class one.
In this scenario, $80\%$ of randomly selected negatively sampled items (class zero) belong to the unpopular category. Therefore, increasing the loss weight for class zero encourages the model to recognize that it is more costly to predict an item as negative when it should be positive.
Formally, the cost-sensitive learning objective in our model is a weighted sum of two one-class classification losses, each computed over positive and negative instances separately.
The new loss function is defined as follows:
\begin{equation}
	Loss_{total} = (1-\lambda) \cdot Loss_{c=1} + (1+ \lambda) \cdot Loss_{c=0}\ ,\label{eq:const-sensitive}
\end{equation}

where $\lambda$ is a hyperparameter that controls the weights of positive and negative classes:
$c=1$ denotes items that the user has interacted with in the past (positive interactions), and $c=0$ refers to negative (uninteracted) items. The class-wise binary cross-entropy terms are defined as:
\begin{align}
	Loss_{c=1} &= - \sum_{i \in I^{+}} \log \tilde{y}_{ui}, \label{eq:loss_pos}\\
	Loss_{c=0} &= - \sum_{j \in I^{-}} \log (1 - \tilde{y}_{uj}), \label{eq:loss_neg}
\end{align}
where $I^{+}$ and $I^{-}$ denote the sets of positively and negatively labeled items, respectively. This design can be interpreted as two one-sided binary classification problems, each targeting a different class, and merged using cost-sensitive coefficients.
In Section~\ref{sec:hyperparameterstudy}, we investigate the role of $\lambda$ in model's performance.
The new loss function can also be represented as a cost matrix, as follows:
\[
\begin{array}{c|c|c}
	& \text{Predicted irrelevant} & \text{Predicted relevant} \\
	\hline
	\text{Actual irrelevant} & 0 & (1-\lambda) Loss_{c=1} \\
	\hline
	\text{Actual relevant} & (1+ \lambda) Loss_{c=0} & 0 \\
\end{array}
\]

\subsection{The algorithm}

In this section, we discuss the details of our two new algorithms, that form the key components of our proposed method: the disentangled popularity and quality algorithm,
and the cost-sensitive edge classification training process.
These algorithms address the issues of popularity bias and misclassification costs in recommendation systems, providing a fairer and more accurate model.

Our approach builds upon the LightGCN architecture for graph-based collaborative filtering, and we adopt the same graph convolution mechanism from LightGCN but make the following modifications: (1) replacement of the BPR loss with the cost-sensitive cross-entropy loss defined in Equation~\ref{eq:const-sensitive}, and (2) addition of the preprocessing step from Algorithm~\ref{alg:disentangled_quality} to filter low-quality edges. All other components—embedding initialization, layer-wise propagation, and training loop structure—remain identical to LightGCN.

\paragraph{Disentangled popularity and quality}
Algorithm~\ref{alg:disentangled_quality} (the disentangled popularity and quality algorithm) aims to filter out low-quality items from the recommendation process by distinguishing between popularity bias and the inherent quality of items.
It consists of the following steps:
\begin{itemize}
	\item 
	First, in Lines $3-10$, we calculate the average rating $\mu$ across all items and compute the $b_u$ and $b_i$ for each user $u$ and item $i$, respectively.
	These deviations are used to estimate the baseline rating $b_{ui}$ for each user-item pair $\left(u,i\right)$.
	\item 
	Then, in Lines $12-17$, the baseline estimation is compared to the actual user rating to calculate the error $e_{ui}$. If an item has a degree of less than $\gamma$ and less than two-thirds of users have given a positive score to that item, the edge between that user and item is removed from the graph,
	effectively filtering out low-quality items that are not wrongly discriminated by popularity bias.
\end{itemize}

\paragraph{Cost-sensitive edge classification training process}
The algorithm for the cost-sensitive edge classification training process (Algorithm~\ref{alg:proposed_method}) addresses the limitations of the BPR loss by replacing it with the negative log-likelihood loss and applying cost-sensitive learning to tackle the misclassification costs associated with popularity bias in recommendation systems.
It consists of the following steps:
\begin{itemize}
	\item 
	First, in Lines $3-6$, the input features are initialized using Xavier initialization, and a Light Graph Convolution is applied to generate node embeddings.
	\item 
	Then, in Line $7$, the algorithm iterates through each user-item pair in the current batch $O_{batch}$.
	For each pair, a negative item is sampled, and the interest scores
	$\tilde{y}_{ui}$
	and $\tilde{y}_{uj}$ are computed for the positive and negative items,
	respectively in Lines $8-10$. 
	\item 
	In Lines $11-12$, instead of the BPR loss, the negative log-likelihood loss is used to calculate the loss for positive interactions $Loss_{c=1}$ and negative interactions $Loss_{c=0}$.
	\item 
	To further enhance fairness, cost-sensitive learning is applied by adjusting the total loss $L_{total}$ based on misclassification costs, controlled by the hyperparameter $\lambda$.
	In Lines $13-19$, the algorithm accumulates the losses
	and updates the model parameters through backpropagation.
	This approach not only replaces the BPR loss with a more robust alternative but also ensures that the model is less biased against unpopular items by accounting for different misclassification costs.
\end{itemize}

\begin{algorithm}[H]
	\caption{The disentangled popularity and quality algorithm.\label{alg:disentangled_quality}}
	\begin{algorithmic}[1]
		\State \textbf{Input:} Graph \(G\left(V,E\right)\), original rating matrix \(R\)
		\State \textbf{Output:} New graph \(G'\left(V,E'\right)\) with filtered edges
		\State Calculate \(\mu \gets\) average rating over all items
		\For{each user \(u\)}
		\State Calculate \(\bar{r}_{u} \gets\) average rating given by user \(u\)
		\State Calculate \(b_u \gets \bar{r}_{u} - \mu\)
		\EndFor
		\For{each item \(i\)}
		\State Calculate \(\bar{r}_{i} \gets\) average rating received by item \(i\)
		\State Calculate \(b_i \gets \bar{r}_{i} - \mu\)
		\EndFor
		\State Initialize new edge set \(E' = E\)
		\For{each edge \((u, i) \in E\)}
		\State Calculate \(b_{ui} \gets \mu + b_u + b_i\) \Comment{Baseline estimation}
		\State Calculate \(e_{ui} \gets r_{ui} - b_{ui}\) \Comment{Calculate error}
		\If{\(\text{degree}(i) < \gamma\) and \( \frac{|\{(u, i) \mid e_{ui} > 0 \}|}{\text{degree}(i)} < \frac{2}{3} \)}
		\State Remove edge \((u, i)\) from \(E'\)
		\EndIf
		\EndFor
		\State \Return new graph \(G'\left(V, E'\right)\)
	\end{algorithmic}
\end{algorithm}

\begin{algorithm}[H]
	\caption{Cost-sensitive edge classification training process.\label{alg:proposed_method}}
	\begin{algorithmic}[1]
		\State \textbf{Input:} Graph \(G\left(V,E\right)\), hyperparameters \(\lambda\) and $\gamma$
		\State \textbf{Output:} Updated model parameters \(\Theta\)
		\State \(X \gets\) Xavier initialization \Comment{Initialize input features}
		\State $H \gets X$
		\State $L \gets 0$
		\State $H \gets Light Graph Convolution()$
		\For{$(u, i) \in O_{batch}$} \Comment{$O_{batch}$ is the set of $\left(u,i\right)$ pairs in the current batch}
		\State Sample a negative item $j$ uniformly at random from $I \setminus O_u^+$
		\State $\tilde{y}_{ui} \gets H[u] \cdot H[i]$ 
		\State $\tilde{y}_{uj} \gets H[u] \cdot H[j]$ 
		
		\State $Loss_{c=1} \gets - \log(\tilde{y}_{ui})$ \Comment{Loss for positive interaction}
		\State $Loss_{c=0} \gets - \log(1 - \tilde{y}_{uj})$ \Comment{Loss for negative interaction}
		
		\State \Comment{Cost-Sensitive Learning}
		\State $Loss_{total} \gets (1 - \lambda)Loss_{c=1} + (1 + \lambda)Loss_{c=0}$
		\State $L \gets L + Loss_{total}$
		\EndFor
		
		\State \textbf{Backpropagate and update} model parameters using gradient descent on $L$
		\State \textbf{return} $\Theta$
	\end{algorithmic}
\end{algorithm}

\paragraph{Time complexity}
We analyze time complexity of our proposed method and compare it to LightGCN, as discussed in \cite{graph-augment}. Time complexity of LightGCN is divided into three main components: graph (adjacency matrix) operations, graph convolution, and loss computation. For graph operations, LightGCN requires \(O(2|E|)\) time to normalize the adjacency matrix. In our proposed method, we additionally compute average ratings, user/item deviations, and filter low-quality items based on their error. This results in a time complexity of \(O(|U| + |I| + |E|)\). The graph convolution stage in LightGCN has a time complexity of \(O(2|E|Kd)\), where \(K\) is the number of layers and $d$ is the embeddings’ size. Our method, which uses Light Graph Convolution, maintains the same time complexity \(O(2|E|Kd)\) during this stage. 
Lastly, for computing the loss function, LightGCN (which employs BPR loss) has a time complexity of \(O(2Bd)\), where \(B\) is the batch size. Our method replaces BPR loss with cross-entropy and incorporates cost-sensitive learning, but its overall time complexity for this component remains \(O(2Bd)\). 
As a result, the whole time complexity of our proposed method is  
\[
O(2|E|Kd + |U| + |I| + 2Bd) = O(|E|Kd + Bd),
\]
which is the same as the whole time complexity of LightGCN. 
Therefore, our fairness‑aware approach does not introduce any extra overhead in terms of time complexity, keeping our model as efficient as LightGCN.

Furthermore, LightGCN’s overall runtime consists exclusively of sparse adjacency normalization, layer‑wise aggregation, and the BPR loss. This combination results in a time complexity that is linear in \(|E|\), \(K\), \(d\), and \(B\), ensuring high efficiency on the GPU. The regularization‑based debiasing model introduces an additional computation—a per‑batch Pearson correlation over \(B\) positive pairs—which incurs a time complexity of \(O(B)\). However, given that \(B \ll |E|K d\), its impact on scalability and throughput is negligible.
Our proposed method requires a one‑time CPU preprocessing cost of \(O(|U|+|I|+|E|)\) to compute user/item statistics and filter edges, while its per‑epoch cost remains \(O(|E|K d + B d)\), thereby matching LightGCN in efficiency and scalability. In contrast, APDA performs, in each of its \(K\) layers, per‑edge cosine similarity (accounting for inverse‑popularity), exponential weight scaling, and per‑node residual updates, leading to a time complexity of \(O((|E|+|U|+|I|) K d)\). Although this complexity is still linear, the additional operations and the requirement to store per‑edge weights substantially increase the constant factors, making APDA the least scalable and efficient among the compared approaches.
Detailed comparisons are provided in Table~\ref{tab:complexity}.

\begin{table}[!h]
	\centering
	\caption{A comparison of computational complexity.}
	\label{tab:complexity}
	\begingroup
	\renewcommand{\arraystretch}{1.2}
	\begin{tabular}{|l|c|c|c|}
		\hline
		\textbf{Method} & \textbf{Time complexity} & \textbf{Additional overhead} & \textbf{Scalability} \\ \hline
		LightGCN~\cite{lightgcn} & \(O(|E|\,K\,d + B\,d)\) & None & High \\ \hline
		Reg~\cite{popularitymetrics} & \(O(|E|\,K\,d + B\,d + B)\) & Per‑batch Pearson correlation \(O(B)\) & High \\ \hline
		Our proposed method & \(O(|E|\,K\,d + B\,d)\) & One‑time filtering \(O(|U|+|I|+|E|)\) & High \\ \hline
		APDA~\cite{apda} & \(O((|E|+|U|+|I|)\,K\,d)\) & Per‑edge cosine + exponential + residuals & Low \\ \hline
	\end{tabular}
	\endgroup
\end{table}
\arrayrulecolor{black}

\section{Experiments}
\label{sec:experiments}

In this section, we explore the effect of the proposed fairness-aware method on both recommendation utility and fairness. We first compare our model with state-of-the-art GNN-based models for a fair recommendation. Subsequently, we study the impact of different components and hyperparameters on the performance of our proposed method\footnote{Our implementation is publicly available at \url{https://github.com/nemat-gholinejad/DisentanglingPopularityQuality}.}.

\subsection{Experimental setup}

\subsubsection{Datasets}

We show the effectiveness of 
our proposed method over five real-world datasets: Bookcrossing\footnote{\url{http://www.bookcrossing.com}}~\cite{book-crossing}, Amazon CDs\footnote{\label{amazondata-footnote}\url{https://cseweb.ucsd.edu/~jmcauley/datasets/amazon/links.html}}~\cite{amazon-data}, Amazon Health\textsuperscript{\ref{amazondata-footnote}}~\cite{amazon-data}, Amazon Toys\textsuperscript{\ref{amazondata-footnote}}~\cite{amazon-data} and Amazon Electronics\textsuperscript{\ref{amazondata-footnote}}~\cite{amazon-data}.
For all datasets, we adopt a $10$-core setting, filtering out users and items with fewer than 
$10$ interactions. We randomly select $70\%$ of each user’s interactions as the training set, $10\%$ for validation, and the last $20\%$ for testing. We use the validation set for hyperparameter tuning and early stopping and report our model performance on the test set as the final result. Table~\ref{tbl:dataset} shows the statistics of the datasets.

During training, we adopt a uniform negative sampling strategy. For each training instance, a user is randomly selected, and one of their interacted items is sampled as a positive example to form a pair \((u, i)\). A negative item \(j\) is then sampled uniformly from the entire item set, excluding items the user has already interacted with (\(I \setminus O_u^+\)). We use one negative sample per positive instance. Negative samples are re-generated at each epoch, ensuring dynamic training pairs and reducing overfitting to fixed negatives.

\begin{table}[!h]
	\centering
	\caption{Dataset statistics.\label{tbl:dataset}}
	\begin{tabular}{@{}l c c c c @{}}
		\toprule
		Dataset & \#user & \#item & \#interaction & Density\\
		\midrule
		Amazon Toys & 921 & 871 & 18637 & 2.32\%\\
		Amazon CDs & 15592 & 16184 & 445412 & 0.176\%\\
		Bookcrossing & 6754 & 13670 & 374325 & 0.4\%\\
		Amazon Electronics & 20242 & 11589 & 347393 & 0.148\%\\
		Amazon Health & 2184 & 1260 & 55076 & 2\%\\
		\bottomrule
	\end{tabular}
\end{table}

\subsubsection{Baselines}

We compare our proposed method against the following baselines:
\begin{itemize}
	\item
	LightGCN~\cite{lightgcn}: This model simplifies graph convolution by removing non-linearity and weight transformation. The final embedding is obtained by averaging the embeddings from all layers.
	
	\item Reg~\cite{popularitymetrics}:
	This method is based on regularization and takes into account the correlation between an item's popularity and its predicted score by the model. As in the original paper, we carefully adjust the hyperparameter $\gamma$ of this model to balance accuracy and fairness in the results.
	
	\item r-AdjNorm~\cite{adjnorm}:
	This model controls the strength of the normalization term to regulate the process during neighborhood aggregation, with a focus on improving results for low-degree items. Based on the paper's guidance, we fine-tune the parameter \(r\) within the range of \([0.5,1.5]\) using a step size of \(0.05\).
	\item
	APDA~\cite{apda}: This method assigns lower weights to connected edges during the aggregation process and uses residual connections to ensure unbiased and fair representations for users and items in graph collaborative filtering. Following the approach in the original paper, we fine-tune the residual parameter $\lambda$ within the range $\left[0, 1.0\right]$.
	
	\item DPNS~\cite{li2025disentangled}:
	This method represents a state-of-the-art progressive negative sampling framework for graph-based collaborative filtering, designed to distinguish between positive and negative user feedback. It projects these two types of feedback into separate low-rank and high-rank eigenvalue spaces, preventing interference during message passing. The model is trained in three phases and incorporates gradient reversal to address the issue of false negatives. Consistent with the original paper, we fine-tune the hyperparameters governing the progressive training process, such as the phase transition parameters  $\beta$ and $\tau$, and the gradient-reversal threshold $T_0$.
\end{itemize}

\subsubsection{Evaluation metrics}

To evaluate the models, we leverage a wide range of evaluation metrics from classical machine learning metrics to ranking metrics as well as fairness metrics.
For utility measurement, we use Recall, NDCG, and MAP (mean average precision). We also utilize three fairness evaluation metrics:
\textit{popularity-rank correlation for users (PRU)}~\cite{popularitymetrics}, \textit{popularity-rank correlation for items (PRI)}~\cite{popularitymetrics}, and \textit{equality of opportunity (EO)}.
In the following, we briefly introduce each of the metrics. Recall is computed as the ratio of relevant items retrieved to the total number of relevant items.
\begin{equation}
	Recall@M = \frac{1}{|U|} \sum_{u \in U} \frac{|\text{Rel}_u(M) \cap \text{GT}_u|}{|\text{GT}_u|},
	\label{eq:recall}
\end{equation}
where $M$ is the number of top recommended items, $GT_u$ is the set of items user $u$ likes during testing, $REL_{u}\left(M\right)$ is the set of top $M$ recommended items for user $u$. NDCG stands for normalized discounted cumulative gain, and 
$NDCG@N$ for top $N$ recommended items is defined as follow:
\begin{equation}
	\small
	NDCG@N = \frac{\sum_{i=1}^{N} \frac{2^{r\left(i\right)}-1}{\log_{2}(i+1)}}{\sum_{i=1}^{REL_N}\frac{2^{r\left(i\right)} - 1}{\log_{2}(i+1)}}\cdot\label{eq:ndcg} 
\end{equation}
Here, \( REL_N \) represents a list of the top \( N \) most relevant items, sorted in ascending order. The function \( r(i) \) indicates the relevance of the item ranked at position \( i \), where \( r(i) \) takes a value of \( 1 \) if the item is relevant, and \( 0 \) if it is irrelevant.
MAP is calculated as the mean of the average precision values across all users:
\begin{equation}
	MAP = \frac{1}{|U|}\sum_{u \in U} \frac{1}{number\ of\ relevant\ documents}\sum_{i=1}^{N}P@i \cdot r(i),\label{eq:map}
\end{equation}
where \(P@i\) is precision at $i$. We next introduce fairness metrics.

\begin{equation}
	PRU = -\frac{1}{|U|} \sum_{u \in U} SRC\left(d_{\tilde{O}_{u}^{+}}, rank_{u}\left(\tilde{O}_{u}^{+}\right)\right)\cdot\label{eq:pru}
\end{equation}
Here, \(SRC\left(\cdot,\cdot\right)\) calculates Spearman's rank correlation, \(\tilde{O}_{u}^{+}\) represents items in the test profile of user $u$, \(rank_{u}\left(x\right)\) signifies the ranking of item $x$ that the model predicts for user $u$ and \(d_{x}\) is the degree of node \(x\).
PRU investigates the correlation between items' popularity and their ranking position
in each user's recommendation list.
This metric ranges between $-1$ and $1$, where $-1$ indicates the model is in its best state considering fairness, and $1$ means unpopular items are placed at the tail of the recommendation list.
$PRI$ is defined as follows:	
\begin{equation}
	PRI = -\sum_{i \in I} SRC\left(d_{i}, \frac{1}{U_{i}}\sum_{u \in U_{i}} rank_{u}\left(i\right)\right)\cdot\label{eq:pri}
\end{equation}
Here \(U_{i}\) is a set of users whose item \(i\) already exists in their test profiles.
PRI analyzes the correlation between items' average position across all users and their popularity. This metric's range is the same as PRU.
$EO$ is defined as follows:
\begin{equation}
	EO = \frac{1}{|U|} \sum_{u \in U} \left| \frac{\sum_{v \in G_0} \mathbf{1}_{v \in \text{TopK}_u \wedge v \in \text{Test}_u} - \sum_{v \in G_1} \mathbf{1}_{v \in \text{TopK}_u \wedge v \in \text{Test}_u}}{\sum_{v \in G_0} \mathbf{1}_{v \in \text{TopK}_u \wedge v \in \text{Test}_u} + \sum_{v \in G_1} \mathbf{1}_{v \in \text{TopK}_u \wedge v \in \text{Test}_u}} \right|,
	\label{eq:eo}
\end{equation}
where $Test_u$ represents the set of items that user $u$ likes during testing.
Larger values for the EO metric are considered indicative of unfairness against unpopular items.

While PRU, PRI, and EO are defined formally above, we further clarify their operational meanings. PRU measures, for each user, whether more popular items systematically receive higher ranking positions. A large positive value indicates that popularity strongly influences ranking, while values closer to $-1$ indicate that ranking is less driven by popularity and therefore fairer toward long-tail items. PRI evaluates this effect from the item perspective by examining whether globally popular items tend to occupy better average ranking positions across users. Similar to PRU, lower values indicate weaker dependence between popularity and ranking. EO measures whether items from two groups (e.g., popular vs. unpopular items) receive similar recommendation accuracy. Operationally, EO captures the difference in true positive rates between the two item groups within users' top-K recommendation lists. An EO value close to $0$ indicates that both groups have similar chances of being correctly recommended, whereas larger values indicate disparity in recommendation quality across groups.
The EO metric is inspired by the classical equality of opportunity definition in fair machine learning, which requires equal true positive rates across sensitive groups. In prior work, EO is often defined over user groups (e.g., demographic groups). In our setting, we adapt this formulation to the item side, where items are divided into two groups based on popularity (e.g., short-head vs. long-tail). This modification aligns with our objective of mitigating popularity bias.
Specifically, instead of comparing true positive rates across user groups, we compare the proportion of correctly recommended items from each item group within users' top-K lists. This adaptation preserves the core principle of equality of opportunity — equal chance of correct recommendation — while making it suitable for item-group fairness analysis. Therefore, although the notation differs slightly from standard user-side EO definitions, the underlying fairness concept remains consistent.

\subsubsection{Initialization of hyperparameters}
To ensure a fair comparison, we adopt uniform training settings for all methods. Specifically, we set the embedding size to 64 and use 3 GNN layers for every model. Following the original implementations, LightGCN, AdjNorm, Reg and Our model use a batch size of 1024, whereas APDA employs a batch size of 2048.
In our model we set $\gamma$ to 20 as its default value, but also explore values such as 15, 25, and 30 in Section~\ref{sec:gamma_effect} for comparative insights.
For $\lambda$, different values between 0.1 and 0.5 are used, partly due to the differing characteristics of these datasets and their respective misclassification penalties.
Additionally, our early-stopping mechanism halts training if NDCG does not improve after 20 epochs. This ensures we first identify the model with optimal accuracy before evaluating its fairness capabilities. This approach emphasizes that a fair model should also maintain accuracy.
Figures~\ref{fig:lambda1} to~\ref{fig:lambda3} show that NDCG remains relatively stable as $\lambda$ moves from 0.0 to 0.4, while fairness metrics exhibit more substantial shifts. This leads us to select 0.1 as the optimal $\lambda$ in some cases.

\subsection{Performance comparison}
In this section, we present a detailed comparison of the performance of our model against the mentioned baselines.
The results are shown for four experimental settings, with evaluations at top-20, top-50, top-100, and top-300 recommendation cutoffs.
We report the results in Tables \ref{tbl:performance_comparison_300} and \ref{tbl:performance_comparison_20}.
Over each dataset and for every metric, the best performance is highlighted in bold, while the second-best is underlined. The key observations are as follows:
\begin{itemize}
	\item
	LightGCN and Reg: These models provide a solid baseline but generally underperform in comparison to state-of-the-art approaches. Their performance is consistent but lower across most datasets and cutoffs. Reg, which incorporates a fairness-based regularization term, shows a slight improvement in ranking metrics. However, it does not significantly enhance the overall recommendation quality.
	On the other hand, considering fairness, these two models perform well on some fairness metrics. Although the results vary across datasets, Reg achieves better fairness than LightGCN in nearly half of the cases.
	This indicates that the added regularization can enhance fairness to some extent.
	However, despite these improvements, the models still struggle with overall fairness. This is evident in other metrics such as EO and PRI, where they exhibit disparities in recommendation quality.
	While they make some strides towards fairness, they lack mechanisms specifically designed to ensure equity across all item groups, which limits their broader fairness impact.
	\item
	r-AdjNorm: This model performs well across all datasets and is generally ranked second or third in performance metrics. r-AdjNorm strikes a good balance between improving the quality of recommendations and maintaining simplicity, but falls behind more complex models in certain cases. In terms of fairness, r-AdjNorm achieves the best metrics in some cases and ranks as the second-best in others, demonstrating strong fairness performance overall—outperforming LightGCN and Reg, but still trailing our proposed method in most scenarios.
	
	\begin{table}[!htb]
		\centering
		\caption{Performance comparison of the models across the used datasets at top-M = 300\label{tbl:performance_comparison_300}}
		\scalebox{0.7}{
			\begin{tabular}{c c c c c c c c | c c c }
				\toprule
				Dataset & Metric &  LightGCN & Reg & r-AdjNorm & APDA & DPNS & Our model$_{2/3}$ & Our model$_{1/2}$ & w/o Cost & w/o Edge \\
				\midrule
				\multirow{5}{*}{Amazon Electronics} & Recall & 0.3017 & 0.2969 & 0.3043 & 0.317 & \underline{0.3327} & \textbf{0.3443} & 0.3038 & 0.3434 & 0.3571 \\
				& NDCG & 0.0937 & 0.0908 & 0.0952 & 0.0991 & \textbf{0.1061} & \underline{0.1006} & 0.993 & 0.0992 & 0.1035\\
				& MAP & 0.0243 & 0.0228 & 0.0254 & \underline{0.0262} & \textbf{0.0294} & \underline{0.0286} & 0.0276 & 0.0274 & 0.0286\\
				\cdashline{2-11}
				& EO & 0.805 & 0.7999 & 0.7697 & \underline{0.7475} & 0.8736 & \textbf{0.6231} & 0.7273 & 0.6159 & 0.6813\\
				& PRU & 0.4864 & \underline{0.4717} & 0.4771 & 0.5254 & 0.5597 & \textbf{0.3148} & 0.3550 & 0.3295 & 0.4431\\
				& PRI & 0.3526 & \underline{0.3445} & 0.3541 & 0.3756 & 0.5462 & \textbf{0.1743} & 0.1726 & 0.2257 & 0.4268\\
				\midrule
				\multirow{5}{*}{Amazon CDs} & Recall & 0.5242 & 0.5319 & 0.5357 & 0.5227 & \underline{0.5487} & \textbf{0.6077} & 0.5563 & 0.6088 & 0.6078\\
				& NDCG & 0.2058 & 0.2064 & \underline{0.2076} & 0.2075 & 0.2005 & \textbf{0.2179} & 0.2084 & 0.2187 & 0.2265 \\
				
				& MAP & 0.0697 & 0.0687 & 0.0691 & \underline{0.0712} & 0.0607 & \textbf{0.0749} & 0.0685 & 0.0754 & 0.0835\\
				\cdashline{2-11}
				& EO & 0.3405 & 0.3549 & 0.3284 & \underline{0.2824} & 0.8249 & \textbf{0.1733} & 0.2633 & 0.1709 & 0.2043\\
				& PRU & \underline{0.2471} & 0.2659 & \textbf{0.2475} & 0.2763 & 0.4678 & 0.2829 & 0.2974 & 0.2782 & 0.2077\\
				& PRI & 0.1482 & 0.1636 & 0.1775 & \underline{0.1627} & 0.4156 & \textbf{0.0214} & -0.0023 & 0.0272 & 0.1849\\
				\midrule
				\multirow{5}{*}{Bookcrossing} & Recall & 0.3185 & 0.314 & \underline{0.334} & 0.322 & 0.3245 & \textbf{0.399} & 0.3769 & 0.4025 & 0.4072\\
				& NDCG & 0.1248 & 0.1238 & \underline{0.1311} & 0.1297 & 0.1273 & \textbf{0.1471} & 0.1400 & 0.1486 & 0.1507\\
				& MAP & 0.0304 & 0.0307 & 0.033 & \underline{0.0331} & 0.0314 & \textbf{0.0414} & 0.0377 & 0.0415 & 0.0418\\
				\cdashline{2-9}
				& EO & 0.6944 & 0.6971 & \underline{0.6292} & 0.6637 & 0.7865 & \textbf{0.4353} & 0.5274 & 0.4577 & 0.5591\\
				& PRU & 0.4121 & 0.4169 & 0.409 & \underline{0.4065} & 0.6342 & \textbf{0.3462} & 0.3954 & 0.359 & 0.3897\\
				& PRI & 0.315 & 0.3198 & 0.3252 & \underline{0.3147} & 0.6383 & \textbf{-0.0033} & 0.1194 & 0.1114 & 0.4145\\
				\midrule
				\multirow{5}{*}{Amazon Health} & Recall & 0.6284 & 0.6246 & \textbf{0.6312}  & \underline{0.6302} & 0.6071 & 0.6182 & 0.6017 & 0.6382 & 0.662 \\
				& NDCG & 0.2356 & 0.2338 & \underline{0.2372} & \textbf{0.2381} & 0.2322 & 0.2316 & 0.2297 & 0.2302 & 0.2446 \\
				& MAP & 0.0715 & 0.0705 & 0.0724  & 0.0734 & 0.0721 & \textbf{0.0764} & 0.0755 & 0.0715 & 0.079 \\
				\cdashline{2-11}
				& EO &  0.2832 & 0.2921 & 0.2874 & \underline{0.2243} & 0.8923 & \textbf{0.1199} & 0.1662 & 0.1548 & 0.239 \\
				& PRU & 0.4131 & \underline{0.4076} & 0.4083 & 0.4226 & 0.5740 & \textbf{0.2001} & 0.2076 & 0.2434 & 0.3796 \\
				& PRI & 0.4088 & 0.3995 & \underline{0.3993} & 0.4207 & 0.5686 & \textbf{0.2427} & 0.2703 & 0.3086 & 0.445 \\
				\midrule
				\multirow{5}{*}{Amazon Toys} & Recall & \underline{0.6034} & 0.6007 & 0.6011 & 0.5994 & 0.5768 & \textbf{0.8113} & 0.6258 & 0.8059 & 0.8121 \\
				& NDCG & \underline{0.1856} & 0.1851 & 0.1838 & 0.1832 & 0.1775 & \textbf{0.2225} & 0.1877 & 0.2220 & 0.2251 \\
				& MAP & 0.0433 & \underline{0.0460} & 0.0420 & 0.0422 & 0.0401 & \textbf{0.058} & 0.0471 & 0.0592 & 0.0588 \\
				\cdashline{2-11}
				& EO & 0.2078  & 0.1964 & 0.1827 & \underline{0.1528} & 0.7298 & \textbf{0.0022} & 0.0459 & 0.0041 & 0.0311 \\
				& PRU & 0.0974 & 0.0908 & \underline{0.0580} & 0.1308 & 0.2803 & \textbf{0.0286} & 0.0149 & 0.0284 & 0.0486 \\
				& PRI & 0.2140 & \underline{0.1465} & 0.1515 & 0.2893 & 0.7045 & \textbf{0.0385} & 0.1192 & 0.0693 & 0.1373 \\
				\bottomrule
			\end{tabular}
		}
	\end{table}

	\begin{table}[!htb]
		\centering
		\caption{Performance comparison of the models across the used datasets at top-M = 100\label{tbl:performance_comparison_100}}
		\scalebox{0.7}{
			\begin{tabular}{c c c c c c c c | c c c }
				\toprule
				Dataset & Metric &  LightGCN & Reg & r-AdjNorm & APDA & DPNS & Our model$_{2/3}$ & Our model$_{1/2}$ & w/o Cost & w/o Edge \\
				\midrule
				\multirow{5}{*}{Amazon Electronics} & Recall & 0.1772 & 0.1726 & 0.1784 & 0.1875 & \underline{0.1987} & \textbf{0.205} & 0.1802 & 0.2011 & 0.2089\\
				& NDCG & 0.0702 & 0.0672 & 0.0712 & 0.0748 & \textbf{0.0810} & \underline{0.0756} & 0.0718 & 0.074 & 0.0773\\
				& MAP & 0.0234 & 0.0218 & 0.0241 & 0.0253 & \textbf{0.0284} & \underline{0.0269} & 0.0248 & 0.0261 & 0.0275\\
				\cdashline{2-11}
				& EO & 0.9119 & 0.9077 & 0.8944 & \underline{0.8674} & 0.9547 & \textbf{0.8493} & 0.8917 & 0.8351 & 0.8535 \\
				& PRU & 0.4793 & 0.4656 & \underline{0.4784} & 0.5225 & 0.5346 & \textbf{0.3185} & 0.3461 & 0.3295 & 0.44\\
				& PRI & 0.348 & \underline{0.3393} & 0.3553 & 0.3739 & 0.5244 & \textbf{0.1944} & 0.1766 & 0.2257 & 0.423\\
				\midrule
				\multirow{5}{*}{Amazon CDs} & Recall & 0.3569 & 0.3567 & 0.3621 & 0.3564 & \underline{0.3595} & \textbf{0.412} & 0.3724 & 0.4155 & 0.417 \\
				& NDCG & 0.1691 & 0.1689 & 0.1697 & \underline{0.1715} & 0.1595 & \textbf{0.1785} & 0. 0.1672 & 0.1798 & 0.1888\\
				& MAP & 0.0667 & 0.0663 & 0.066 & \underline{0.0689} & 0.0574 & \textbf{0.0718} & 0.0668 & 0.0722 & 0.0811\\
				\cdashline{2-11}
				& EO & 0.5069 & 0.5222 & 0.4761 & \underline{0.4439} & 0.8452 & \textbf{0.3656} & 0.4243 & 0.368 & 0.358\\
				& PRU & \textbf{0.2471} & 0.2558 & \underline{0.2475} & 0.2746 & 0.4707 & 0.2844 & 0.2997 & 0.2771 & 0.1987\\
				& PRI & \underline{0.1482} & 0.1616 & 0.1775 & 0.1601 & 0.4221 & \textbf{0.0001} & -0.0017 & 0.026 & 0.1838\\
				\midrule
				\multirow{5}{*}{Bookcrossing} & Recall & 0.1862 & 0.1851 & \underline{0.1964} & 0.1881 & 0.1919 & \textbf{0.2423} & 0.2275 & 0.2432 & 0.2409\\
				& NDCG & 0.0922 & 0.0916 & 0.0974 & \underline{0.0975} & 0.0957 & \textbf{0.1121} & 0.1060 & 0.1128 & 0.113\\
				& MAP & 0.0280 & 0.0282 & 0.0307 & \underline{0.0316} & 0.0294 & \textbf{0.0391} & 0.0354 & 0.0393 & 0.0393\\
				\cdashline{2-11}
				& EO & 0.7707 & 0.7714 & 0.7166 & \underline{0.7132} & 0.8432 & \textbf{0.5884} & 0.6515 & 0.6131 & 0.676\\
				& PRU & 0.4158 & 0.4091 & 0.407 & \underline{0.3786} & 0.6271 & \textbf{0.3469} & 0.3954 & 0.351 & 0.3796\\
				& PRI & 0.3173 & 0.3195 & 0.3278 & \underline{0.3005} & 0.6308 & \textbf{-0.0041} & 0.0690 & 0.1096 & 0.409\\
				\midrule
				\multirow{5}{*}{Amazon Health} 
				& Recall & 0.3732 & 0.3697 & \textbf{0.3773} & 0.3741 & 0.3721 & \underline{0.3743} & 0.3622 & 0.3676 & 0.3944 \\
				& NDCG & 0.1809 & 0.1788 & \underline{0.1825} & \textbf{0.1833} & 0.182 & 0.1808 & 0.1792 & 0.1728 & 0.1885 \\
				& MAP & 0.0672 & 0.0659 & 0.0677 & \underline{0.0693} & 0.0682 & \textbf{0.0726} & 0.0702 & 0.0663 & 0.0744 \\
				\cdashline{2-11}
				& EO & 0.5527 & 0.5655 & 0.5435 & \underline{0.531} & 0.8788 & \textbf{0.3732} & 0.4238 & 0.4286 & 0.5221 \\
				& PRU & \underline{0.4033} & 0.4255 & 0.4144 & 0.4113 & 0.53 & \textbf{0.2005} & 0.2104 & 0.2716 & 0.3746 \\
				& PRI & \underline{0.4005} & 0.4066 & 0.4016 & 0.4149 & 0.5247 & \textbf{0.2447} & 0.2852 & 0.3427 & 0.4412 \\
				\midrule
				\multirow{5}{*}{Amazon Toys} 
				& Recall & \underline{0.3126} & 0.3021 & 0.2958 & 0.3143 & 0.2993 & \textbf{0.4347} & 0.3332 & 0.4085 & 0.4379 \\
				& NDCG & \underline{0.1250} & 0.1212 & 0.1611 & 0.1175 & 0.1195 & \textbf{0.1553} & 0.1302 & 0.1491 & 0.1579 \\
				& MAP & \underline{0.0386} & 0.0371 & 0.0363 & 0.0366 & 0.0372 & \textbf{0.0542} & 0.0435 & 0.0518 & 0.0547 \\
				\cdashline{2-11}
				& EO & 0.4813 & 0.4559 & \underline{0.4011} & 0.4691 & 0.7496 & \textbf{0.1696} & 0.2748 & 0.2635 & 0.2136 \\
				& PRU & 0.0979 & \underline{0.0821} & 0.1451 & 0.1283 & 0.4124 & \textbf{0.0063} & 0.0042 & 0.0169 & 0.0273 \\
				& PRI & 0.2124 & \underline{0.1029} & 0.1343 & 0.2745 & 0.8636 & \textbf{0.0283} & 0.1305 & 0.0953 & 0.0911 \\
				\bottomrule
			\end{tabular}
		}
	\end{table}

	\begin{table}[!htb]
		\centering
		\caption{Performance comparison of the models across the used datasets at top-M =50\label{tbl:performance_comparison_50}}
		\scalebox{0.7}{
			\begin{tabular}{c c c c c c c c | c c c }
				\toprule
				Dataset & Metric &  LightGCN & Reg & r-AdjNorm & APDA & DPNS & Our model$_{2/3}$ & Our model$_{1/2}$ & w/o Cost & w/o Edge \\
				\midrule
				\multirow{5}{*}{Amazon Electronics} 
				& Recall & 0.1216 & 0.1178 & 0.1228 & 0.1286 & \underline{0.1394} & \textbf{0.1404} & 0.1254 & 0.1377 & 0.143 \\
				& NDCG & 0.0575 & 0.0552 & 0.0584 & 0.0616 & \textbf{0.0673} & \underline{0.0624} & 0.0589 & 0.0604 & 0.0631 \\
				& MAP & 0.0223 & 0.0212 & 0.0231 & 0.0245 & \textbf{0.0272} & \underline{0.0264} & 0.0252 & 0.025 & 0.0261\\
				\cdashline{2-11}
				& EO & 0.9344 & 0.9355 & 0.9304 & \textbf{0.8963} & 0.9624 & \underline{0.9294} & 0.9314 & 0.9124 & 0.9192\\
				& PRU & 0.4679 & \underline{0.457} & 0.4784 & 0.5061 & 0.5222 & \textbf{0.3081} & 0.3447 & 0.3311 & 0.4463\\
				& PRI & 0.3391 & \underline{0.3334} & 0.3553 & 0.3572 & 0.5108 & \textbf{0.1595} & 0.1675 & 0.228 & 0.4286\\
				\midrule
				\multirow{5}{*}{Amazon CDs} 
				& Recall & 0.2679 & 0.2675 & 0.2679 & \underline{0.267} & 0.2579 & \textbf{0.3075} & 0.2719 & 0.3083 & 0.3135 \\
				& NDCG & 0.1469 & 0.1461 & \underline{0.1436} & 0.1486 & 0.1337 & \textbf{0.1543} & 0.1426 & 0.1544 & 0.1641 \\
				& MAP & 0.0642 & 0.0634 & 0.0609 & \underline{0.0659} & 0.0541 & \textbf{0.0691} & 0.0613 & 0.069 & 0.0777\\
				\cdashline{2-11}
				& EO & 0.6023 & 0.6175 & 0.5487 & \underline{0.5405} & 0.8278 & \textbf{0.5064} & 0.5781 & 0.5148 & 0.4605\\
				& PRU & \underline{0.2477} & 0.2571 & \textbf{0.2442} & 0.2717 & 0.4679 & 0.2714 & 0.2852 & 0.2773 & 0.1987\\
				& PRI & \underline{0.1489} & 0.161 & 0.1734 & 0.1585 & 0.4167 & \textbf{0.0107} & -0.0165 & 0.0256 & 0.1838\\
				\midrule
				\multirow{5}{*}{Bookcrossing} 
				& Recall & 0.1269 & 0.1235 & \underline{0.1342} & 0.1304 & 0.131 & \textbf{0.1711} & 0.1625 & 0.169 & 0.1701\\
				& NDCG & 0.0759 & 0.0752 & 0.0794 & \underline{0.0818} & 0.0793 & \textbf{0.094} & 0.0884 & 0.0934 & 0.0943\\
				& MAP & 0.0262 & 0.027 & 0.0283 & \underline{0.0302} & 0.0272 & \textbf{0.0375} & 0.0337 & 0.0374 & 0.037\\
				\cdashline{2-11}
				& EO & 0.7783 & 0.7713 & \underline{0.7013} & 0.7283 & 0.8761 & \textbf{0.6549} & 0.6978 & 0.6634 & 0.7093\\
				& PRU & 0.4188 & 0.4059 & 0.4093 & \underline{0.3744} & 0.6268 & \textbf{0.3482} & 0.3874 & 0.353 & 0.3806\\
				& PRI & 0.3167 & 0.315 & 0.3198 & \underline{0.2916} & 0.6314 & \textbf{0.053} & 0.0632 & 0.111 & 0.4099\\
				\midrule
				\multirow{5}{*}{Amazon Health} 
				& Recall & 0.2605 & 0.2601 & 0.2647 & \underline{0.2669} & 0.2583 & \textbf{0.2679} & 0.2590 & 0.2598 & 0.2809 \\
				& NDCG & 0.1493 & 0.1496 & 0.1525 & \textbf{0.1545} & 0.1510 & \underline{0.1541} & 0.1514 & 0.1443 & 0.1588 \\
				& MAP & 0.0619  & 0.062 & 0.0635 & \underline{0.0652} & 0.0634 & \textbf{0.0695} & 0.0669 & 0.0624 & 0.0702 \\
				\cdashline{2-11}
				& EO & 0.7478  & 0.7254 & \underline{0.6664} & 0.6913 & 0.8713 & \textbf{0.5402} & 0.6059 & 0.6219 & 0.6989 \\
				& PRU & 0.4517 & 0.4442 & \underline{0.4083} & 0.4113 & 0.5269 & \textbf{0.1995} & 0.2250 & 0.3 & 0.3796 \\
				& PRI & 0.4198 & 0.4157 & \underline{0.3993} & 0.4149 & 0.5174 & \textbf{0.241} & 0.2922 & 0.3671 & 0.445 \\
				\midrule
				\multirow{5}{*}{Amazon Toys} 
				& Recall & 0.2057 & 0.2017 & 0.2058 & \underline{0.2224} & 0.2135 & \textbf{0.2777} & 0.2159 & 0.2684 & 0.2641 \\
				& NDCG & 0.0985 & 0.0971 & 0.0987 & \underline{0.1047} & 0.1023 & \textbf{0.1206} & 0.1025 & 0.1181 & 0.1161 \\
				& MAP & 0.0360 & 0.0356 & 0.0364 & \underline{0.0388} & 0.0370 & \textbf{0.0496} & 0.0408 & 0.0488 & 0.0478 \\
				\cdashline{2-11}
				& EO & 0.6264 & 0.5953 & \underline{0.4783} & 0.5472 & 0.8443 & \textbf{0.2806} & 0.4062 & 0.3511 & 0.3601 \\
				& PRU & 0.0979 & 0.0941 & \underline{0.0874} & 0.0881 & 0.1973 & \textbf{0.0286} & 0.0090 & 0.0974 & 0.0941 \\
				& PRI & 0.2124 & 0.1854 & \underline{0.1352} & 0.2972 & 0.7218 & \textbf{0.0385} & 0.1184 & 0.1276 & 0.1473 \\
				\bottomrule
			\end{tabular}
		}
	\end{table}
	
	\begin{table}[!htb]
		\centering
		\caption{Performance comparison of the models across the used datasets at top-M =20\label{tbl:performance_comparison_20}}
		\scalebox{0.7}{
			\begin{tabular}{c c c c c c c c | c c c }
				\toprule
				Dataset & Metric &  LightGCN & Reg & r-AdjNorm & APDA & DPNS & Our model$_{2/3}$ & Our model$_{1/2}$ & w/o Cost & w/o Edge \\
				\midrule
				\multirow{5}{*}{Amazon Electronics} 
				& Recall & 0.0721 & 0.0678 & 0.0716 & 0.0764 & \underline{0.0823} & \textbf{0.0837} & 0.0788 & 0.0818 & 0.0844 \\
				& NDCG & 0.0438 & 0.0415 & 0.0444 & 0.0467 & \textbf{ 0.0515} & \underline{0.0477} & 0.0451 & 0.0459 & 0.0478\\
				& MAP & 0.0205 & 0.0195 & 0.0212 & 0.0221 & \textbf{0.0251} & \underline{0.0243} & 0.0219 & 0.0228 & 0.0239 \\
				\cdashline{2-11} 
				& EO & 0.9618 & 0.9502 & \underline{0.9472} & \textbf{0.924} & 0.9758 & 0.9648 & 0.9710 & 0.9527 & 0.9547 \\
				& PRU & 0.4696 & \underline{0.4393} & 0.4711 & 0.5015 & 0.5260 & \textbf{0.3081} & 0.3396 & 0.3295 & 0.4431 \\
				& PRI & 0.3409 & \underline{0.3149} & 0.3518 & 0.3531 & 0.5140 & \textbf{0.1595} & 0.1616 & 0.2257 & 0.4268 \\
				\midrule
				\multirow{5}{*}{Amazon CDs} & Recall & 0.1678 & 0.1709 & 0.171 & \underline{0.1713} & 0.1549 & \textbf{0.1904} & 0.1667 & 0.1923 & 0.2051 \\
				& NDCG & 0.1165 & 0.1169 & 0.1173 & \underline{0.12} & 0.1033 & \textbf{0.121} & 0.1095 & 0.1222 & 0.1337 \\
				& MAP & 0.0574 & 0.0569 & 0.057 & \underline{0.0601} & 0.0480 & \textbf{0.0619} & 0.0534 & 0.0625 & 0.0711 \\
				\cdashline{2-11}
				& EO & 0.7131 & 0.7196 & \underline{0.6481} & \textbf{0.6432} & 0.8324 & 0.6707 & 0.7142 & 0.6712 & 0.5782 \\
				& PRU & \underline{0.2474} & 0.2608 & \textbf{0.239} & 0.2665 & 0.4674 & 0.285 & 0.322 & 0.2781 & 0.1952 \\
				& PRI & \underline{0.1491} & 0.1632 & 0.1792 & 0.1538 & 0.4146 & \textbf{0.0006} & -0.0027 & 0.0251 & 0.1839 \\
				\midrule
				\multirow{5}{*}{Bookcrossing} & Recall & 0.0754 & 0.0736 & 0.0791 & \underline{0.0792} & 0.0787 & \textbf{0.1038} & 0.0971 & 0.1043 & 0.1017 \\
				& NDCG & 0.0615 & 0.0603 & 0.0644 & \underline{0.0673} & 0.0637 & \textbf{0.0749} & 0.0696 & 0.0749 & 0.0754 \\
				& MAP & 0.0236 & 0.0239 & 0.0261 & \underline{0.0271} & 0.0242 & \textbf{0.0339} & 0.0301 & 0.0338 & 0.0334 \\
				\cdashline{2-11}
				& EO & 0.7565 & 0.7441 & \textbf{0.6875} & 0.7324 & 0.8475 & \underline{0.6909} & 0.7192 & 0.6961 & 0.7204 \\
				& PRU & 0.4121 & 0.3931 & 0.4012 & \underline{0.3637} & 0.6224 & \textbf{0.3462} & 0.3701 & 0.3529 & 0.3796 \\
				& PRI & 0.3150 & 0.3125 & 0.3276 & \underline{0.2896} & 0.6222 & \textbf{0.0511} & 0.0690 & 0.113 & 0.409 \\
				\midrule
				\multirow{5}{*}{Amazon Health} 
				& Recall & 0.1597 & 0.1595 & 0.159 & \underline{0.1644} & 0.1596 & \textbf{0.1781} & 0.1736 & 0.1649 & 0.1798 \\
				& NDCG & 0.1176 & 0.117 & 0.1183 & \underline{0.1212} & 0.1199 & \textbf{0.1246} & 0.1239 & 0.1143 & 0.127 \\
				& MAP & 0.0553 & 0.0547 & 0.0559 & \underline{0.0575} & 0.0571 & \textbf{0.0624} & 0.0618 & 0.0562 & 0.0633 \\
				\cdashline{2-11}
				& EO & \underline{0.8252} & 0.8357 & 0.7749 & 0.8271 & 0.92 & \textbf{0.6849} & 0.7014 & 0.7324 & 0.8064 \\
				& PRU & 0.4188  & 0.4442 & \underline{0.4083} & 0.4116 & 0.5179 & \textbf{0.2001} & 0.2076 & 0.3012 & 0.3796 \\
				& PRI & 0.4095 & 0.4154 & \underline{0.3993}  & 0.4144 & 0.5076 & \textbf{0.2427} & 0.2730 & 0.3692 & 0.445 \\
				\midrule
				\multirow{5}{*}{Amazon Toys} & Recall & \underline{0.1094} & 0.1072 & 0.0986 & 0.1053 & 0.1028 & \textbf{0.1531} & 0.1174 & 0.1341 & 0.1553 \\
				& NDCG & 0.0682 & 0.0671 & \underline{0.0687} & 0.0623 & 0.0643 & \textbf{0.0868} & 0.0732 & 0.0798 & 0.0884 \\
				& MAP & \underline{0.0306} & 0.0291 & 0.0283 & 0.0301 & 0.0295 & \textbf{0.0436} & 0.0353 & 0.0408 & 0.0428 \\
				\cdashline{2-11}
				& EO & 0.7464 & 0.7081 & \underline{0.6401} & 0.6702 & 0.7923 & \textbf{0.4473} & 0.5600 & 0.5495 & 0.5674 \\
				& PRU & 0.0804 & \underline{0.0784} & 0.0829 & 0.0783 & 0.1093 & \textbf{0.0286} & 0.0115 & 0.1185 & 0.0954 \\
				& PRI & 0.2089 & 0.1835 & \underline{0.1318} & 0.2629 & 0.2351 & \textbf{0.0385} & 0.1145 & 0.1101 & 0.1314 \\
				\bottomrule
			\end{tabular}
		}
	\end{table}

	\item
	APDA: This model remains the most powerful baseline overall. While it is no longer the top performer in all scenarios, APDA consistently ranks second in most cases across both performance and fairness metrics.
	In terms of fairness, APDA also demonstrates solid performance—ranking second in most cases and even outperforming all other methods in certain instances. This makes APDA a strong candidate in settings where both accuracy and fairness are important, though our proposed method still offers more balanced improvements.
	\item 
	DPNS: In evaluating this model, it can be observed that its performance quality is highly dependent on the dataset. For instance, it achieves the best results on the Electronics dataset.
	However, from a fairness perspective, the model performs poorly, consistently yielding the lowest fairness scores across all datasets.
	A closer look at its behavior on the Electronics dataset reveals that, although it achieves high accuracy, it suffers from severe fairness issues. This highlights the importance of maintaining a reasonable balance between accuracy and fairness.
	
	The poor fairness performance of DPNS can be explained by its objective and design. The method is primarily designed to improve ranking performance by mitigating false negatives and generating harder negative samples within the BPR framework. Its progressive negative sampling strategy tends to focus on “hard” negatives, which are often popular items that are genuinely irrelevant to the user. While this improves the model’s discriminative ability, it does not explicitly address popularity bias. As a result, long-tail items remain under-sampled, and the model further reinforces the dominance of popular items in the embedding space. Consequently, DPNS learns to rank popular items more aggressively, achieving high accuracy but failing to provide balanced exposure across item groups, which leads to poor fairness performance.
	
	\item
	Our model (2/3): Our model with the default filtering threshold (two-thirds positive feedback) remains the strongest overall approach. It achieves the top or second-best performance in most cases across all datasets in terms of accuracy metrics such as Recall, NDCG, and MAP. In terms of fairness, it ranks first in nearly all cases and remains highly competitive elsewhere, demonstrating its ability to deliver both accurate and equitable recommendations. Moreover, in the few cases where its accuracy is slightly lower than another model, a comparison of the corresponding fairness metrics shows a remarkable improvement in fairness despite the minor drop in accuracy. This balance makes our model particularly well-suited for scenarios where both performance and fairness are essential, as it promotes a more balanced treatment of item groups without disproportionately favoring or disadvantaging any. We note that this setting corresponds to our default threshold; the effect of alternative thresholds is analyzed in Section~\ref{sec:threshold_effect}.
\end{itemize}

\begin{table}[!htb]
	\centering
	\caption{t-test results comparing our model with competing methods across accuracy and fairness metrics.}
	\label{tbl:ttest_results}
	\begin{tabular}{lccccc}
		\toprule
		Metric & Ours vs.\ DPNS & Ours vs.\ APDA & Ours vs.\ r-AdjNorm & Ours vs.\ Reg & Ours vs.\ LightGCN \\
		\midrule
		Recall & $\checkmark$ & $\checkmark$ & $\checkmark$ & $\checkmark$ & $\checkmark$ \\
		NDCG   & $\checkmark$ & $\checkmark$ & $\checkmark$ & $\checkmark$ & $\checkmark$ \\
		MAP    & --              & --              & --              & --              & -- \\
		EO     & $\checkmark$ & $\checkmark$ & $\checkmark$ & $\checkmark$ & $\checkmark$ \\
		PRU    & $\checkmark$ & $\checkmark$ & $\checkmark$ & $\checkmark$ & $\checkmark$ \\
		PRI    & $\checkmark$ & $\checkmark$ & $\checkmark$ & $\checkmark$ & $\checkmark$ \\
		\bottomrule
	\end{tabular}
\end{table}

\subsection{Significance analysis of our improvements over baseline models}

In this section, we examine the statistical significance of the improvements achieved by our proposed model compared to the competing methods. To this end, we conduct a paired t-test with the standard threshold of $p = 0.05$. The test is applied to all results (top-$M = 300, 100, 50, 20$) of each competing model against our method, across each accuracy and fairness evaluation metric. The outcomes are summarized in Table~\ref{tbl:ttest_results}.
In this table, the symbol $\checkmark$ indicates that our model significantly outperforms the corresponding competing method under the t-test, meaning that the observed improvement is statistically significant. The symbol {\bf --} denotes that the improvement is not statistically significant.	
As shown in the table, for all fairness metrics the improvements achieved by our proposed method over every competing model are statistically significant. Furthermore, for two of the three accuracy metrics our method shows statistically significant improvements over all competitors; for the MAP metric the improvement does not reach statistical significance.

We note that accuracy improvements often trade off against fairness: methods that increase accuracy frequently perform poorly on fairness metrics, and vice versa. Therefore, an appropriate goal when developing fair recommendation methods is to substantially improve fairness without causing a meaningful degradation in accuracy. Our proposed model achieves this objective by substantially improving all fairness metrics while also providing statistically significant improvements for most accuracy metrics.

\subsection{Ablation study}

In this section, we analyze the influence of individual components of our proposed method on the final output. Specifically, we evaluate the effects of our Edge-classification and Cost-sensitive learning techniques on the model. These scenarios are denoted as “w/o Edge” and “w/o Cost”, respectively. The corresponding results are presented in the last two columns of Tables~\ref{tbl:performance_comparison_300} to~\ref{tbl:performance_comparison_20}. Note that, since our cost-sensitive weighting scheme cannot be applied to the BPR loss function, it is omitted in the “w/o Edge” variant.

Considering recommendation accuracy, removing the cost-sensitive component from our model generally results in a slight decline in performance--especially on the Amazon CDs and Bookcrossing datasets. The "w/o Cost" variant also shows a significant drop in fairness, whereas our complete model achieves improved EO, PRU, and PRI scores, highlighting the importance of incorporating fairness mechanisms.
The "w/o Edge" variant represents the LightGCN model applied to our cleaned dataset with our disentangled popularity and quality method. A similar pattern emerges as with the "w/o Cost" variant, though the differences are more pronounced. Specifically, while the "w/o Edge" variant exhibits a modest improvement in accuracy, its fairness metrics decline considerably. This observation implies that although the BPR loss function can enhance accuracy, it tends to favor users interested in popular items, thus compromising fairness and limiting the model's ability to serve users with less affinity for popular items.

To better understand the relative impact of each component on fairness, we analyze why removing edge classification leads to a larger degradation than removing cost-sensitive learning in some cases.
This behavior can be explained by the different roles of the two components. In datasets with strong popularity skew (e.g., Amazon Electronics and BookCrossing), the "w/o Edge" variant—where BPR is retained—exhibits significantly worse fairness (higher EO, PRU, and PRI) compared to "w/o Cost". This is because the BPR objective introduces a systematic bias by encouraging all observed interactions to be ranked above unobserved ones, which tends to favor popular items that have more interactions.
In contrast, the edge classification framework replaces this pairwise ranking objective with a pointwise formulation, allowing the model to evaluate each interaction independently. This reduces the implicit pressure to suppress long-tail items, enabling the model to assign more balanced scores across items with different popularity levels. Cost-sensitive learning further refines this behavior by reweighting misclassification penalties, but it operates on top of the underlying objective.
Therefore, removing edge classification reintroduces the primary source of popularity bias, leading to a larger degradation in fairness, while removing cost-sensitive learning mainly reduces an additional corrective effect. This also explains why the gap between "w/o Edge" and "w/o Cost" is smaller in less skewed datasets (e.g., Amazon Toys), where the impact of popularity bias is less pronounced.

\subsection{Hyperparameters study}
\label{sec:hyperparameterstudy}

\subsubsection{Effect of $\lambda$}
\label{sec:hyperparameter_lambda_effect}

\begin{figure}[H]
	\begin{subfigure}[b]{0.49\linewidth}
		\includegraphics[width=\linewidth]{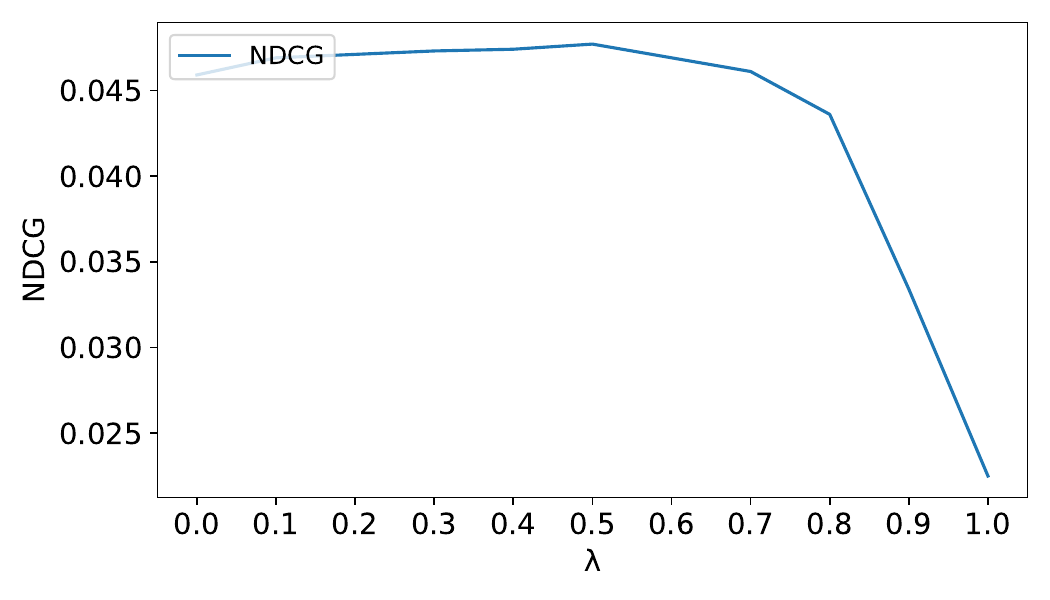}
		\caption{}
	\end{subfigure}
	\begin{subfigure}[b]{0.49\linewidth}
		\includegraphics[width=\linewidth]{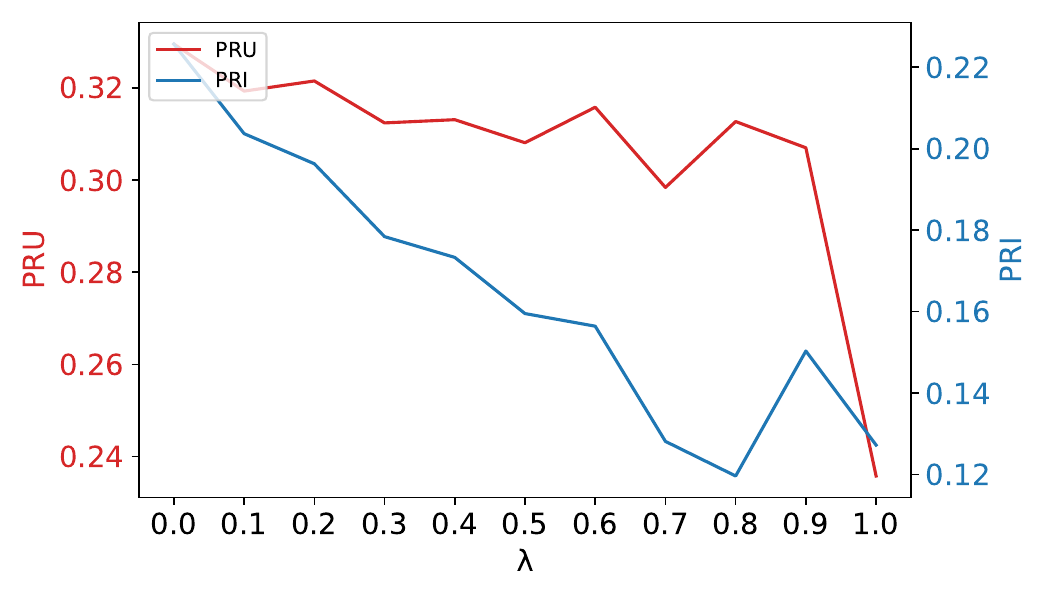}
		\caption{}
	\end{subfigure}
	\caption{Studying the effect of $\lambda$, over the Electronics dataset,
		while setting $\gamma$ to $20$. \label{fig:lambda1}}
\end{figure}

\begin{figure}[H]
	\begin{subfigure}[b]{0.49\linewidth}
		\includegraphics[width=\linewidth]{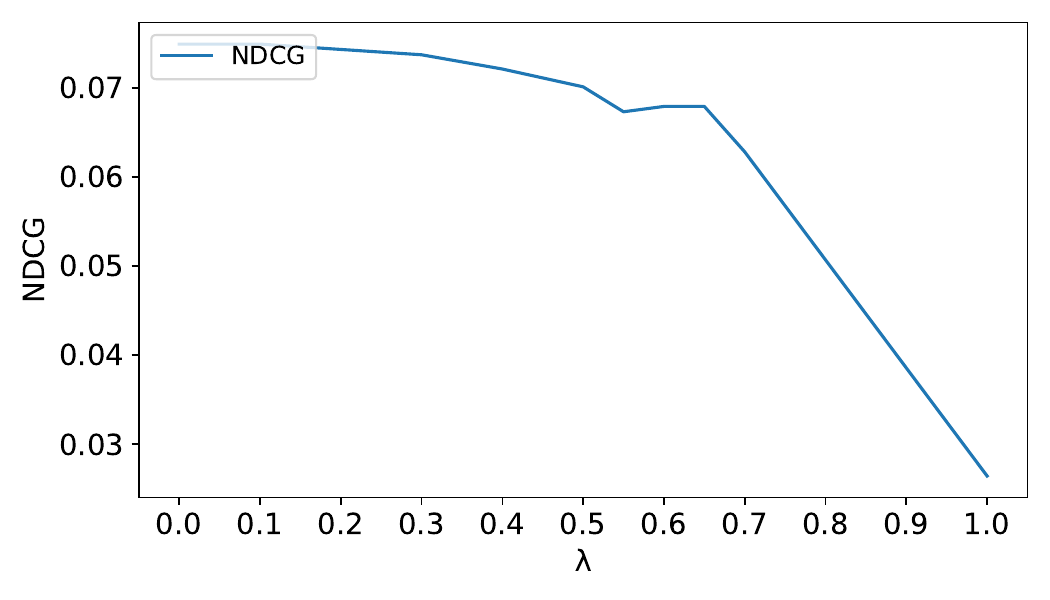}
		\caption{}
	\end{subfigure}
	\begin{subfigure}[b]{0.49\linewidth}
		\includegraphics[width=\linewidth]{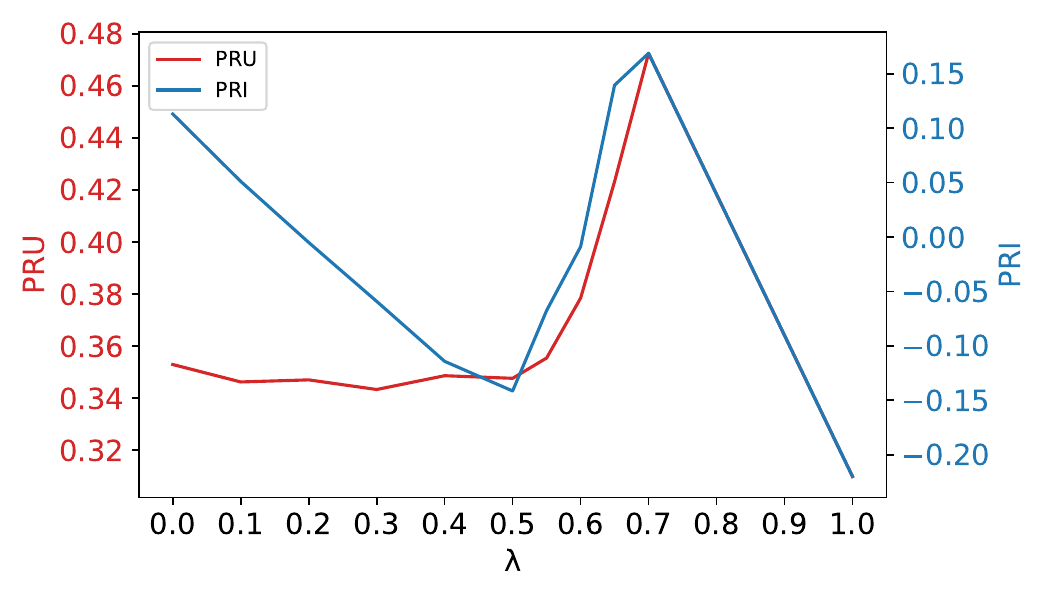}
		\caption{}
	\end{subfigure}
	\caption{Studying the effect of $\lambda$, over the bookcrossing dataset, while setting $\gamma$ to $20$. \label{fig:lambda2}}
\end{figure}

\begin{figure}[H]
	\begin{subfigure}[b]{0.49\linewidth}
		\includegraphics[width=\linewidth]{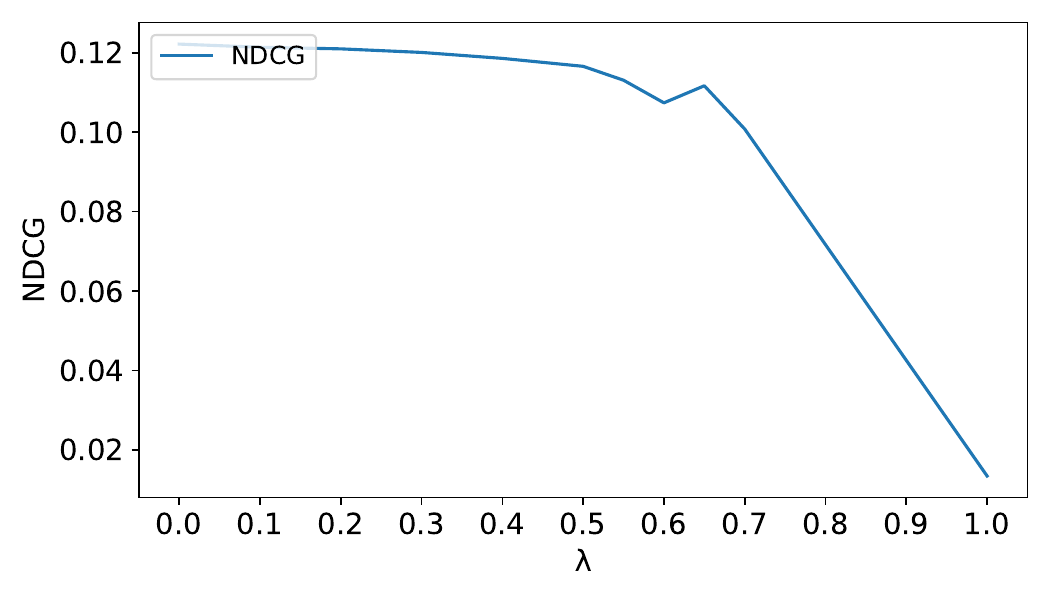}
		\caption{}
	\end{subfigure}
	\begin{subfigure}[b]{0.49\linewidth}
		\includegraphics[width=\linewidth]{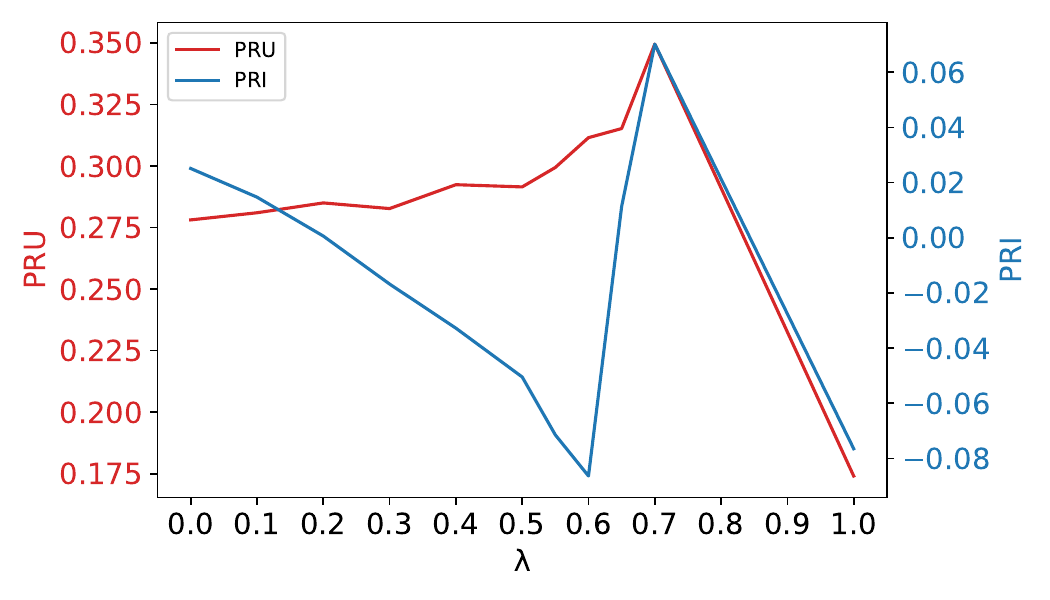}
		\caption{}
	\end{subfigure}
	\caption{Studying the effect of $\lambda$, over the CDs dataset, while setting $\gamma$ to $20$. \label{fig:lambda3}}
\end{figure}

In this section, we analyze the impact of tuning the cost-sensitive learning hyperparameter $\lambda$ on recommendation accuracy and fairness across three datasets: Electronics, Bookcrossing, and CDs. We evaluate the model's behavior by varying $\lambda$ between 0 and 1 in increments of 0.1, using metrics such as PRU, PRI, and NDCG. The results are shown in Figures \ref{fig:lambda1} to \ref{fig:lambda3}.
Across all datasets, we observe a consistent and interpretable trend: as the cost-sensitive learning parameter $\lambda$ increases from 0, both PRU and PRI metrics initially decline. This reduction indicates improved fairness, as the model begins to prioritize under-recommended, less popular items more effectively. In this early phase, small to moderate values of $\lambda$ enable the model to rebalance exposure by correcting the dominance of highly popular items, a common source of unfairness in recommender systems. This is particularly important in long-tail domains, where excessive popularity bias can prevent valuable niche items from being surfaced to users.
However, this improvement in fairness does not persist indefinitely. As $\lambda$ continues to rise beyond a moderate threshold, typically around $0.5$ to $0.6$, we begin to observe saturation in fairness gains. In some cases, both PRU and PRI even show signs of reversal, suggesting that the model starts to overcorrect. This turning point reflects the onset of diminishing returns: the cost penalty imposed on popular item misclassifications becomes disproportionately large, nudging the model to suppress popular items more than is optimal. As a result, the fairness advantage stalls and may regress due to the distortion of natural item relevance.
Simultaneously, we observe a clear downward trajectory in NDCG as $\lambda$ increases. While mild reductions in NDCG are tolerable and often expected in fairness-oriented models, the accuracy loss becomes pronounced at higher values of $\lambda$. This underscores a trade-off in fairness-aware learning: stronger fairness constraints can inadvertently undermine the model's capacity to accurately rank the most relevant items, which are often popular. At high $\lambda$ values, the model sacrifices too much relevance in pursuit of balance, leading to degraded user experience in practical deployment.
This pattern yields a practical takeaway for hyperparameter selection. $\lambda$ is indeed a powerful lever for improving fairness, especially in the early range. Moderately sized values of $\lambda$ (e.g., in the range $0.3$–$0.6$) generally offer a desirable balance, providing significant fairness improvements with only minor losses in accuracy. Pushing $\lambda$ too far, on the other hand, results in overcompensation, whereby the model's ability to discern quality recommendations is compromised.
So, these findings emphasize the importance of controlled fairness interventions. Fairness goals and dataset characteristics should be considered when tuning this parameter. The analysis highlights that while tuning $\lambda$ can improve fairness, it must be done cautiously to prevent a significant loss of accuracy in the recommendation system.

Based on these observations, we further analyze how $\lambda$ interacts with dataset characteristics. The effect of $\lambda$ is closely related to the imbalance and sparsity of the dataset. In highly skewed, sparse datasets like Amazon Electronics, a larger portion of items lies in the long tail, and moderate $\lambda$ values help recover under-exposed but relevant items, leading to substantial fairness improvements with minimal impact on accuracy. In contrast, for denser datasets such as Amazon Toys, smaller $\lambda$ values are preferable, as overly large $\lambda$ may overpenalize negative samples and suppress genuinely relevant popular items. This behavior is consistent with Figures~\ref{fig:lambda1} to \ref{fig:lambda3}, where fairness improves steadily with increasing $\lambda$, but NDCG begins to decline earlier in denser datasets due to overcorrection. Therefore, $\lambda$ should be tuned based on dataset sparsity, selecting the largest value that improves fairness while maintaining acceptable accuracy.

\subsubsection{Effect of $\gamma$}
\label{sec:gamma_effect}

\begin{figure}[H]
	\begin{subfigure}[b]{0.49\linewidth}
		\includegraphics[width=\linewidth]{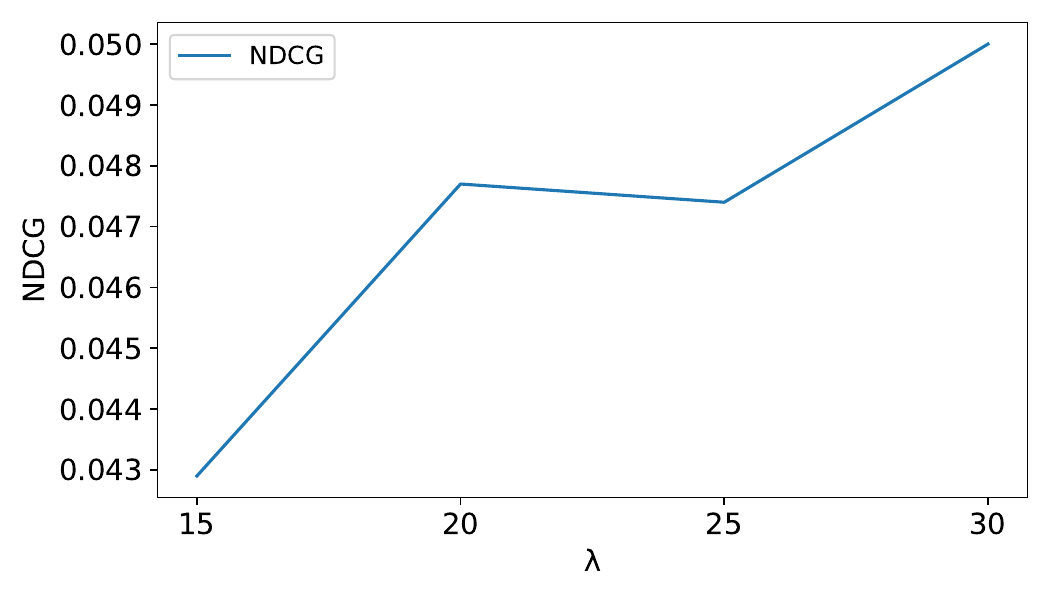}
		\caption{}
	\end{subfigure}
	\begin{subfigure}[b]{0.49\linewidth}
		\includegraphics[width=\linewidth]{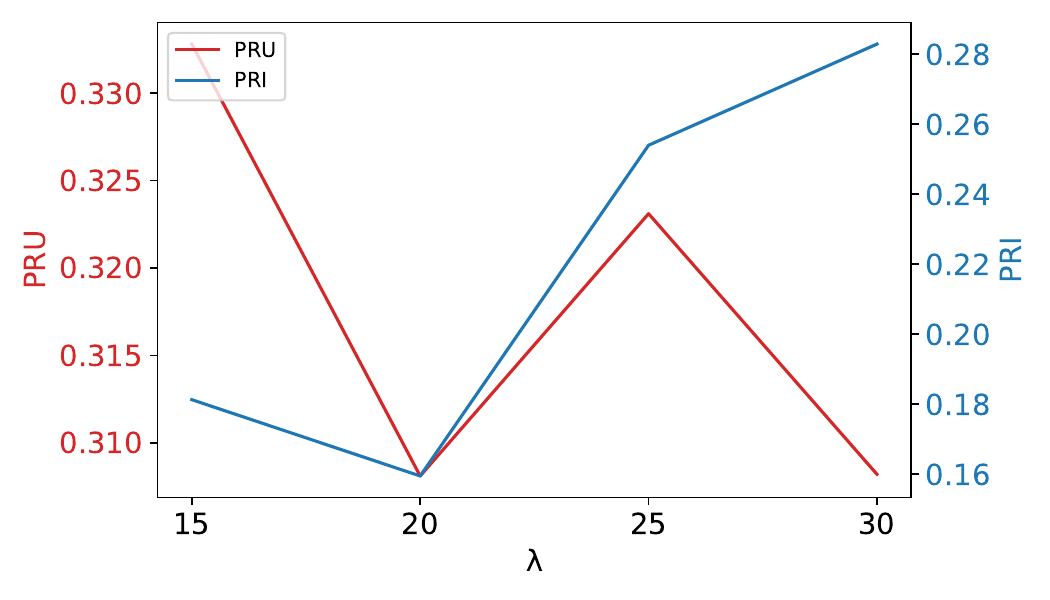}
		\caption{}
	\end{subfigure}
	\caption{Studying the effect of $\gamma$ over the Electronics dataset. \label{fig:gamma1}}
\end{figure}

\begin{figure}[H]
	\begin{subfigure}[b]{0.49\linewidth}
		\includegraphics[width=\linewidth]{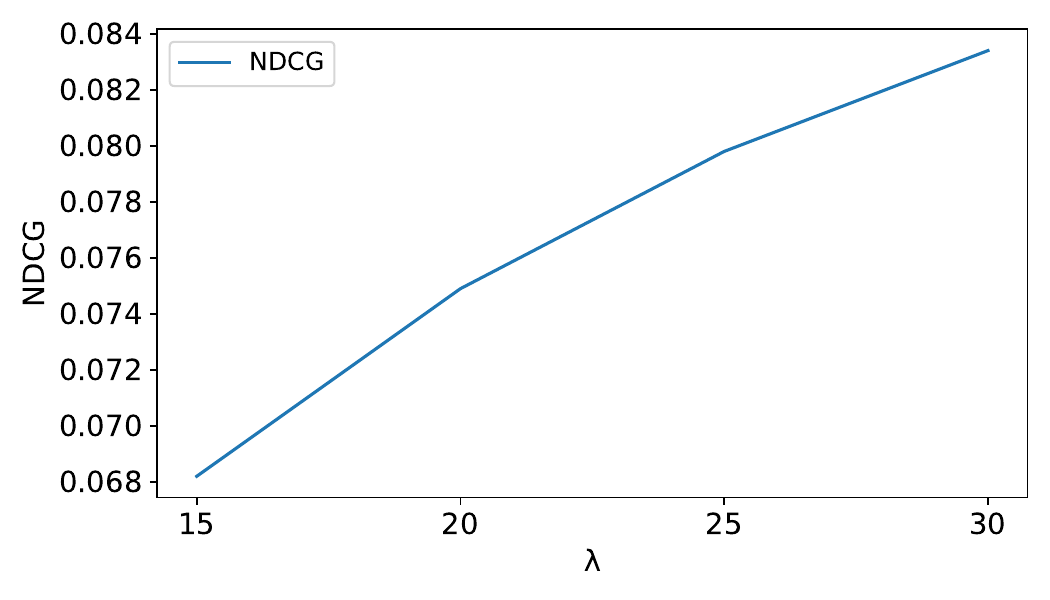}
		\caption{}
	\end{subfigure}
	\begin{subfigure}[b]{0.49\linewidth}
		\includegraphics[width=\linewidth]{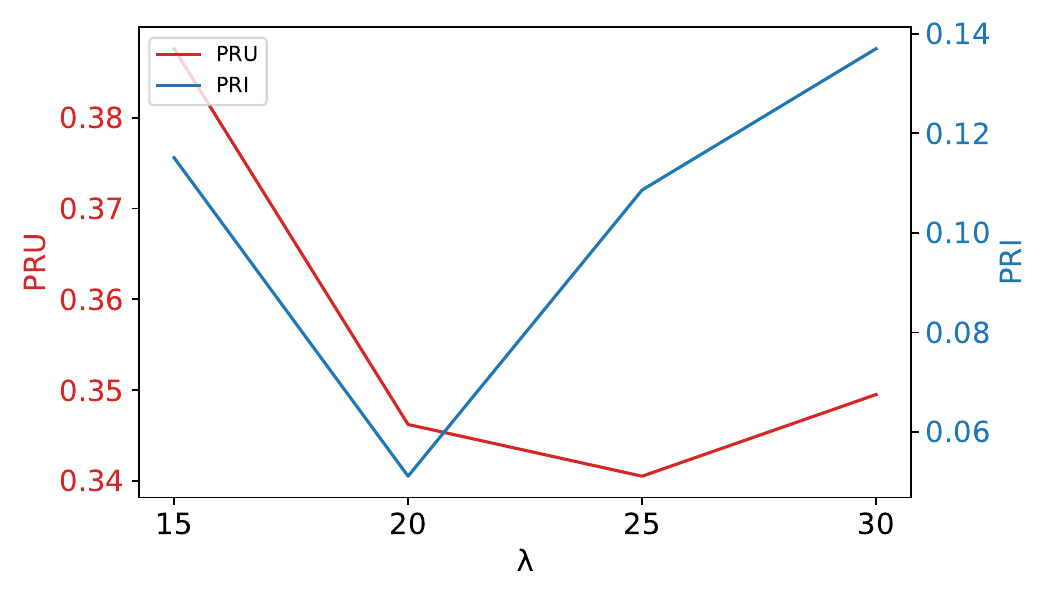}
		\caption{}
	\end{subfigure}
	\caption{Studying the effect of $\gamma$, over the bookcrossing dataset. \label{fig:gamma2}}
\end{figure}

\begin{figure}[H]
	\begin{subfigure}[b]{0.49\linewidth}
		\includegraphics[width=\linewidth]{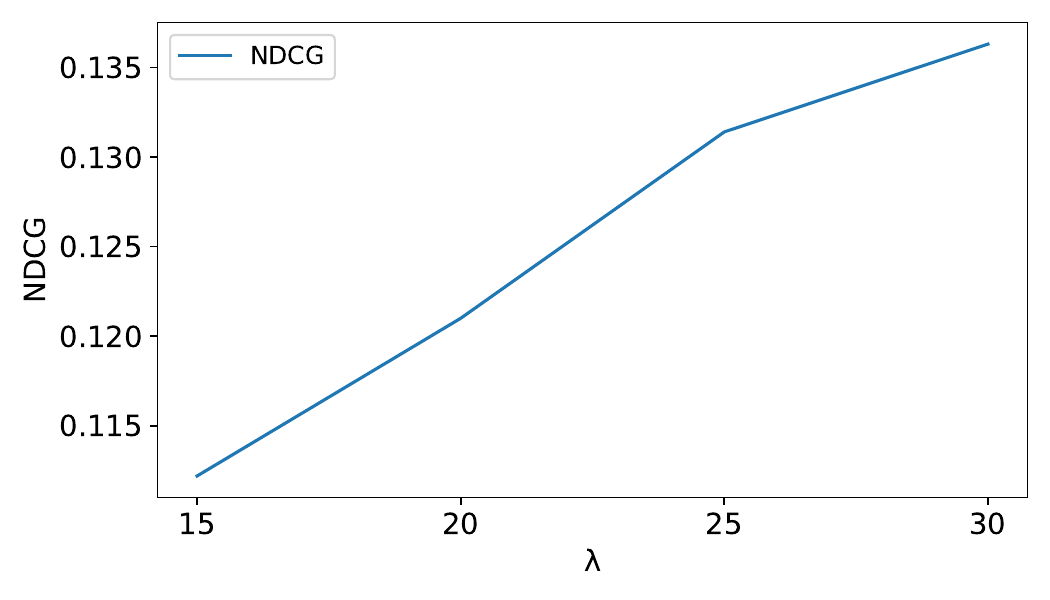}
		\caption{}
	\end{subfigure}
	\begin{subfigure}[b]{0.49\linewidth}
		\includegraphics[width=\linewidth]{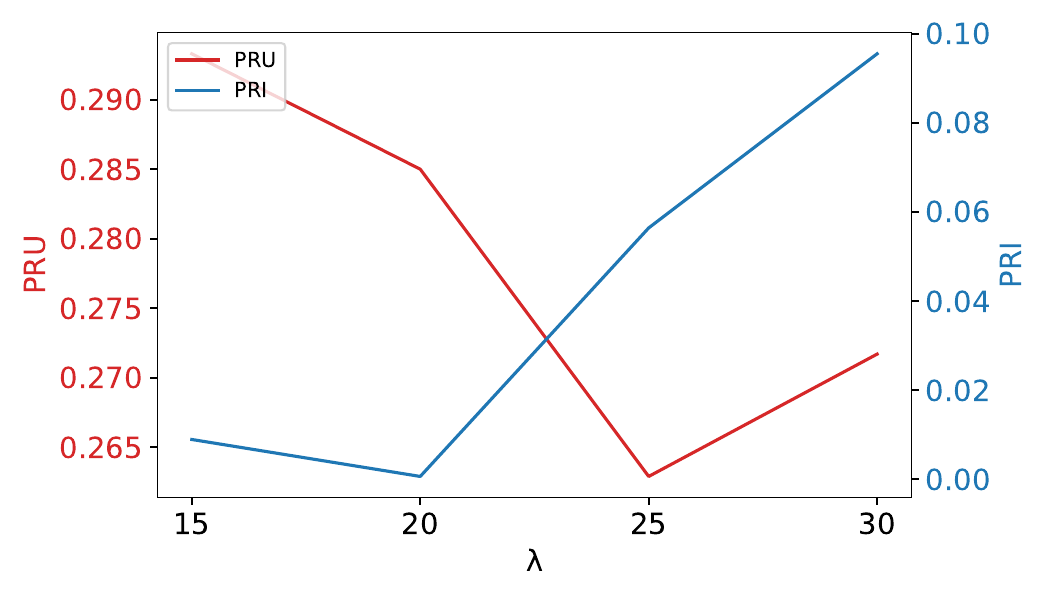}
		\caption{}
	\end{subfigure}
	\caption{Studying the effect of $\gamma$, over the CDs dataset. \label{fig:gamma3}}
\end{figure}

In this section, we analyze the impact of adjusting the hyperparameter $\gamma$ in intervals from $15$ to $30$, with steps of $5$. This parameter study investigates how $\gamma$ influences recommendation accuracy and fairness, focusing on three metrics: PRU, PRI, and NDCG.
The results are depicted in Figures \ref{fig:gamma1} to \ref{fig:gamma3}. We observe that as $\gamma$ grows, the model consistently achieves improvements in NDCG. This consistent upward trend in accuracy suggests that larger values of $\gamma$ enhance the model’s ability to leverage a broader range of user-item interactions.
However, the fairness perspective reveals a more nuanced pattern. As $\gamma$ increases, PRU tends to decline across datasets, reflecting a shift toward fairer treatment of less popular items. This reduction implies that, even while accuracy improves, the model does not disproportionately favor popular items. In effect, increasing $\gamma$ seems to encourage a fairer distribution of recommendation opportunities across the item spectrum, without significantly penalizing recommendation quality. This outcome challenges the common assumption that accuracy and fairness must always be in strict opposition.
The behavior of PRI under changing $\gamma$ values reveals another layer of control. Initially, PRI values dip, indicating a more equitable exposure of item popularity groups. However, beyond a certain point, PRI begins to rise again, suggesting a subtle return of imbalance. This behavior may reflect a tipping point in which the broader inclusion of interaction data starts to reintroduce popularity bias.
These shifts indicate that adjusting $\gamma$ allows for finer control over fairness, balancing exposure for both popular and less popular items without sacrificing the overall accuracy.

\subsubsection{Effect of filtering threshold}
\label{sec:threshold_effect}
To assess the robustness of the proposed filtering strategy, we compare the default threshold (two-thirds positive feedback) with a more relaxed setting (one-half) across all datasets.
The results show that changing the threshold does not significantly degrade model performance, indicating that our framework is stable with respect to this design choice. Across all datasets, the 1/2 threshold leads to only a minor decrease in accuracy metrics while the overall ranking quality remains comparable to the default setting.
However, a consistent pattern emerges in fairness metrics: the relaxed threshold results in slightly worse fairness,higher EO and PRU, across most datasets. This behavior can be explained by the nature of the additional edges retained under the 1/2 setting. Specifically, lowering the threshold allows more interactions to remain in the graph, but a considerable portion of these correspond to low-quality or ambiguous feedback, where user preference is weak or inconsistent. As a result, the model is exposed to noisier supervision signals, which reduces its ability to correctly identify and promote genuinely relevant long-tail items.
From a technical perspective, the filtering step acts as a denoising mechanism: the stricter 2/3 threshold removes edges that are unlikely to reflect true user preference, leading to cleaner gradients during training and more reliable separation between positive and negative signals. In contrast, the 1/2 threshold weakens this effect by introducing label noise, which particularly affects fairness metrics that depend on correctly identifying underexposed but relevant items.
Importantly, the fact that fairness degrades only moderately—and accuracy remains relatively stable,demonstrates that the proposed method is robust to reasonable variations of the threshold. At the same time, the consistent advantage of the 2/3 setting suggests that stricter filtering provides a better trade-off by effectively removing unreliable interactions while preserving meaningful long-tail signals.

\subsection{Model Interpretability}
\label{sec:interpretability}

In this section, we analyze our model from the interpretability perspective. Understanding why a model makes particular recommendations is crucial for both transparency and user trust, especially when balancing fairness and accuracy. Because our framework integrates edge classification and cost-sensitive learning within a GNN, every recommendation can be traced back to clear, human-understandable components.
\paragraph{Edge Classification}  
As described in Section~\ref{sec:edgeclassification}, we first classify each edge in the user-item graph \(G=(V,E)\) to separate genuine preference signals from popularity bias. This classification immediately explains recommendation inclusion or exclusion:
\begin{itemize}
	\item  Recommended long-tail items, even those with a low degree, are those whose edges survived filtering by surpassing the quality threshold.  
	\item Excluded popular items are those whose edges were deemed low-quality, signaling that their popularity did not reflect true user interest.
\end{itemize}
By inspecting the retained edges for a user \(u\), one can see exactly which past interactions drove the recommendation. In particular, the classification confidence derived from the error term \(e_{ui}=r_{ui}-b_{ui}\) indicates whether the model’s decision was quality-driven or bias-filtered.
\paragraph{Cost-Sensitive Weighting}  
The cost-sensitive component further enhances interpretability by assigning asymmetric penalties \((1-\lambda)\) and \((1+\lambda)\) to false positives and false negatives, respectively. From an interpretability standpoint:
\begin{itemize}
	\item When a less popular item is recommended, the higher penalty on false negatives has tipped the balance.  
	\item Conversely, if an item is not recommended despite positive edge classification, the weighting reveals that avoiding an irrelevant recommendation was deemed more important.
\end{itemize}
Thus, every final score can be decomposed into two contributions, edge-based quality signals and class-weight adjustments, making the model’s internal reasoning fully transparent. Reporting \(\lambda\) alongside recommendations clarifies how strongly fairness was enforced in a given run.
Together, these mechanisms ensure that our GNN-based recommender is not only accurate and fair but also inherently interpretable.

\section{Limitations and Future Work}
\label{sec:limitations}

While our method demonstrates strong empirical performance, its limitations should also be acknowledged:

\begin{itemize}
	\item
	\textbf{Dependence on explicit ratings:} 
	The quality filtering step (Algorithm~\ref{alg:disentangled_quality}) relies on explicit numerical ratings to compute baseline estimates. While this assumption holds for many rating-based datasets, its direct application to purely implicit-feedback settings (e.g., clicks, views, purchases) is not straightforward. In such cases, the baseline estimator can be adapted by treating interactions as binary signals and incorporating confidence weights (e.g., interaction frequency or dwell time) as proxies for preference strength. Alternatively, the baseline component could be replaced with a popularity-adjusted confidence model or a learned relevance estimator. Exploring such adaptations is an important direction for future work.
	
	\item
	\textbf{Computational overhead of filtering:} 
	Although the per-epoch training complexity is comparable to LightGCN, the one-time preprocessing step—scanning all edges to compute baseline estimates—can be computationally expensive for very large-scale graphs (e.g., billions of interactions). In such settings, approximate filtering strategies, such as sampling-based methods or incremental updates, would be required to maintain scalability.
\end{itemize}

Beyond addressing these limitations, we plan to: (i) extend the framework to incorporate multi-modal information (e.g., item images or textual descriptions), (ii) investigate the robustness of the model under adversarial popularity manipulation, and (iii) use network centrality indices \cite{DBLP:conf/cikm/ChehreghaniBA19} for item quality estimation.

\section{Conclusion}
\label{sec:conclusion}

In this paper, we proposed a novel approach that tackles exposure bias in recommendation systems through an edge classification framework. Our method reclassifies the edges within the user-item interaction graph, distinguishing between high-quality and low-quality long-tail items. This ensures that popularity bias is mitigated without sacrificing the recommendation of genuinely relevant items. Furthermore, it uses cost-sensitive learning to adjust the misclassification penalties, particularly for unpopular items.
Our experiments on several well-known datasets demonstrated that our proposed model significantly outperforms state-of-the-art methods in terms of fairness metrics. Moreover, it achieves accuracy results very close to those of the best existing methods.


\section*{Acknowledgment}

This work is supported by the Iran National Science Foundation (INSF) under project No. 4034377.

\section*{Data availability statement}
The datasets used in this study (Bookcrossing~\cite{book-crossing}, Amazon CDs~\cite{amazon-data}, Amazon Health~\cite{amazon-data}, Amazon Toys~\cite{amazon-data} and Amazon Electronics~\cite{amazon-data}) are publicly available from their respective providers. 
The code used to generate the experimental results is publicly available at \url{https://github.com/nemat-gholinejad/DisentanglingPopularityQuality}.

\bibliographystyle{plain}	
\bibliography{cite}

@article{DBLP:journals/tjs/ZohrabiSC24,
  author       = {Mohammadjavad Zohrabi and
                  Saeed Saravani and
                  Mostafa Haghir Chehreghani},
  title        = {Centrality-based and similarity-based neighborhood extension in graph
                  neural networks},
  journal      = {J. Supercomput.},
  volume       = {80},
  number       = {16},
  pages        = {24638--24663},
  year         = {2024},
  url          = {https://doi.org/10.1007/s11227-024-06336-x},
  doi          = {10.1007/S11227-024-06336-X},
  bibsource    = {dblp computer science bibliography, https://dblp.org}
}

@inproceedings{DBLP:conf/cikm/ChehreghaniBA19,
  author       = {Mostafa Haghir Chehreghani and
                  Albert Bifet and
                  Talel Abdessalem},
  editor       = {Wenwu Zhu and
                  Dacheng Tao and
                  Xueqi Cheng and
                  Peng Cui and
                  Elke A. Rundensteiner and
                  David Carmel and
                  Qi He and
                  Jeffrey Xu Yu},
  title        = {Adaptive Algorithms for Estimating Betweenness and \emph{k}-path Centralities},
  booktitle    = {Proceedings of the 28th {ACM} International Conference on Information
                  and Knowledge Management, {CIKM} 2019, Beijing, China, November 3-7,
                  2019},
  pages        = {1231--1240},
  publisher    = {{ACM}},
  year         = {2019},
  url          = {https://doi.org/10.1145/3357384.3358064},
  doi          = {10.1145/3357384.3358064},
  bibsource    = {dblp computer science bibliography, https://dblp.org}
}

@inproceedings{elkan2001foundations,
	title={The foundations of cost-sensitive learning},
	author={Elkan, Charles},
	booktitle={International joint conference on artificial intelligence},
	volume={17},
	number={1},
	pages={973--978},
	year={2001},
	organization={Lawrence Erlbaum Associates Ltd}
}

@article{DBLP:journals/natmi/Chehreghani22,
  author       = {Mostafa Haghir Chehreghani},
  title        = {Half a decade of graph convolutional networks},
  journal      = {Nat. Mach. Intell.},
  volume       = {4},
  number       = {3},
  pages        = {192--193},
  year         = {2022},
  url          = {https://doi.org/10.1038/s42256-022-00466-8},
  doi          = {10.1038/S42256-022-00466-8},
  bibsource    = {dblp computer science bibliography, https://dblp.org}
}

@article{10.1145/3700790,
  author       = {Fatemeh Gholamzadeh Nasrabadi and
                  AmirHossein Kashani and
                  Pegah Zahedi and
                  Mostafa Haghir Chehreghani},
  title        = {Content Augmented Graph Neural Networks},
  journal      = {{ACM} Trans. Web},
  volume       = {19},
  number       = {4},
  pages        = {40:1--40:19},
  year         = {2025},
  url          = {https://doi.org/10.1145/3700790},
  doi          = {10.1145/3700790},
  bibsource    = {dblp computer science bibliography, https://dblp.org}
}

@inproceedings{ijcnncost,
	title={Cost-sensitive learning methods for imbalanced data},
	author={Thai-Nghe, Nguyen and Gantner, Zeno and Schmidt-Thieme, Lars},
	booktitle={The 2010 International joint conference on neural networks (IJCNN)},
	pages={1--8},
	year={2010},
	organization={IEEE}
}

@inproceedings{nipsunbiased,
	title={Unbiased pairwise learning from implicit feedback},
	author={Saito, Yuta},
	booktitle={NeurIPS 2019 Workshop on Causal Machine Learning},
	year={2019}
}

@inproceedings{apda,
	title={Adaptive popularity debiasing aggregator for graph collaborative filtering},
	author={Zhou, Huachi and Chen, Hao and Dong, Junnan and Zha, Daochen and Zhou, Chuang and Huang, Xiao},
	booktitle={Proceedings of the 46th International ACM SIGIR Conference on Research and Development in Information Retrieval},
	pages={7--17},
	year={2023}
}

@inproceedings{graph-augment,
	title={Are graph augmentations necessary? simple graph contrastive learning for recommendation},
	author={Yu, Junliang and Yin, Hongzhi and Xia, Xin and Chen, Tong and Cui, Lizhen and Nguyen, Quoc Viet Hung},
	booktitle={Proceedings of the 45th international ACM SIGIR conference on research and development in information retrieval},
	pages={1294--1303},
	year={2022}
}

@inproceedings{amazon-data,
	title={Ups and downs: Modeling the visual evolution of fashion trends with one-class collaborative filtering},
	author={He, Ruining and McAuley, Julian},
	booktitle={proceedings of the 25th international conference on world wide web},
	pages={507--517},
	year={2016}
}

@article{gcmc,
	title={Graph convolutional matrix completion},
	author={Berg, Rianne van den and Kipf, Thomas N and Welling, Max},
	journal={arXiv preprint arXiv:1706.02263},
	year={2017}
}

@inproceedings{wu2019simplifying,
	title={Simplifying graph convolutional networks},
	author={Wu, Felix and Souza, Amauri and Zhang, Tianyi and Fifty, Christopher and Yu, Tao and Weinberger, Kilian},
	booktitle={International conference on machine learning},
	pages={6861--6871},
	year={2019},
	organization={PMLR}
}

@inproceedings{lrgccf,
	title={Revisiting graph based collaborative filtering: A linear residual graph convolutional network approach},
	author={Chen, Lei and Wu, Le and Hong, Richang and Zhang, Kun and Wang, Meng},
	booktitle={AAAI conference on artificial intelligence},
	volume={34},
	number={01},
	pages={27--34},
	year={2020}
}

@inproceedings{sgl,
	title={Self-supervised graph learning for recommendation},
	author={Wu, Jiancan and Wang, Xiang and Feng, Fuli and He, Xiangnan and Chen, Liang and Lian, Jianxun and Xie, Xing},
	booktitle={44th international ACM SIGIR conference on research and development in information retrieval},
	pages={726--735},
	year={2021}
}

@inproceedings{popularitymetrics,
	title={Popularity-opportunity bias in collaborative filtering},
	author={Zhu, Ziwei and He, Yun and Zhao, Xing and Zhang, Yin and Wang, Jianling and Caverlee, James},
	booktitle={Proceedings of the 14th ACM International Conference on Web Search and Data Mining},
	pages={85--93},
	year={2021}
}

@inproceedings{book-crossing,
	title={Improving recommendation lists through topic diversification},
	author={Ziegler, Cai-Nicolas and McNee, Sean M and Konstan, Joseph A and Lausen, Georg},
	booktitle={Proceedings of the 14th international conference on World Wide Web},
	pages={22--32},
	year={2005}
}

@inproceedings{lightgcn,
	title={Lightgcn: Simplifying and powering graph convolution network for recommendation},
	author={He, Xiangnan and Deng, Kuan and Wang, Xiang and Li, Yan and Zhang, Yongdong and Wang, Meng},
	booktitle={Proceedings of the 43rd International ACM SIGIR conference on research and development in Information Retrieval},
	pages={639--648},
	year={2020}
}

@inproceedings{adjnorm,
	title={Investigating accuracy-novelty performance for graph-based collaborative filtering},
	author={Zhao, Minghao and Wu, Le and Liang, Yile and Chen, Lei and Zhang, Jian and Deng, Qilin and Wang, Kai and Shen, Xudong and Lv, Tangjie and Wu, Runze},
	booktitle={45th International ACM SIGIR Conference on Research and Development in Information Retrieval},
	pages={50--59},
	year={2022}
}

@article{bpr,
	title={BPR: Bayesian personalized ranking from implicit feedback},
	author={Rendle, Steffen and Freudenthaler, Christoph and Gantner, Zeno and Schmidt-Thieme, Lars},
	journal={arXiv preprint arXiv:1205.2618},
	year={2012}
}

@article{gcn,
	title={Semi-supervised classification with graph convolutional networks},
	author={Kipf, Thomas N and Welling, Max},
	journal={arXiv:1609.02907},
	year={2016}
}

@inproceedings{ngcf,
	title={Neural graph collaborative filtering},
	author={Wang, Xiang and He, Xiangnan and Wang, Meng and Feng, Fuli and Chua, Tat-Seng},
	booktitle={Proceedings of the 42nd international ACM SIGIR conference on Research and development in Information Retrieval},
	pages={165--174},
	year={2019}
}

@inproceedings{dgrec,
	title={DGRec: Graph Neural Network for Recommendation with Diversified Embedding Generation},
	author={Yang, Liangwei and Wang, Shengjie and Tao, Yunzhe and Sun, Jiankai and Liu, Xiaolong and Yu, Philip S and Wang, Taiqing},
	booktitle={Proceedings of the Sixteenth ACM International Conference on Web Search and Data Mining},
	pages={661--669},
	year={2023}
}

@inproceedings{peng2024powerful,
	title={How powerful is graph filtering for recommendation},
	author={Peng, Shaowen and Liu, Xin and Sugiyama, Kazunari and Mine, Tsunenori},
	booktitle={Proceedings of the 30th ACM SIGKDD Conference on Knowledge Discovery and Data Mining},
	pages={2388--2399},
	year={2024}
}

@article{peng2024less,
	title={Less is More: Removing Redundancy of Graph Convolutional Networks for Recommendation},
	author={Peng, Shaowen and Sugiyama, Kazunari and Mine, Tsunenori},
	journal={ACM Transactions on Information Systems},
	volume={42},
	number={3},
	pages={1--26},
	year={2024},
	publisher={ACM New York, NY}
}

@inproceedings{rastegarpanah2019fighting,
	title={Fighting fire with fire: Using antidote data to improve polarization and fairness of recommender systems},
	author={Rastegarpanah, Bashir and Gummadi, Krishna P and Crovella, Mark},
	booktitle={Proceedings of the twelfth ACM international conference on web search and data mining},
	pages={231--239},
	year={2019}
}

@inproceedings{ekstrand2018all,
	title={All the cool kids, how do they fit in?: Popularity and demographic biases in recommender evaluation and effectiveness},
	author={Ekstrand, Michael D and Tian, Mucun and Azpiazu, Ion Madrazo and Ekstrand, Jennifer D and Anuyah, Oghenemaro and McNeill, David and Pera, Maria Soledad},
	booktitle={Conference on fairness, accountability and transparency},
	pages={172--186},
	year={2018},
	organization={PMLR}
}

@inproceedings{damak2021debiased,
	title={Debiased explainable pairwise ranking from implicit feedback},
	author={Damak, Khalil and Khenissi, Sami and Nasraoui, Olfa},
	booktitle={Proceedings of the 15th ACM Conference on Recommender Systems},
	pages={321--331},
	year={2021}
}

@inproceedings{gtn,
	title={Graph trend filtering networks for recommendation},
	author={Fan, Wenqi and Liu, Xiaorui and Jin, Wei and Zhao, Xiangyu and Tang, Jiliang and Li, Qing},
	booktitle={Proceedings of the 45th international ACM SIGIR conference on research and development in information retrieval},
	pages={112--121},
	year={2022}
}

@inproceedings{anelli2023auditing,
	title={Auditing consumer-and producer-fairness in graph collaborative filtering},
	author={Anelli, Vito Walter and Deldjoo, Yashar and Di Noia, Tommaso and Malitesta, Daniele and Paparella, Vincenzo and Pomo, Claudio},
	booktitle={European Conference on Information Retrieval},
	pages={33--48},
	year={2023},
	organization={Springer}
}

@article{zhao2022popularity,
	title={Popularity bias is not always evil: Disentangling benign and harmful bias for recommendation},
	author={Zhao, Zihao and Chen, Jiawei and Zhou, Sheng and He, Xiangnan and Cao, Xuezhi and Zhang, Fuzheng and Wu, Wei},
	journal={IEEE Transactions on Knowledge and Data Engineering},
	year={2022},
	publisher={IEEE}
}

@article{chen2024graph,
	title={How graph convolutions amplify popularity bias for recommendation?},
	author={Chen, Jiajia and Wu, Jiancan and Chen, Jiawei and Xin, Xin and Li, Yong and He, Xiangnan},
	journal={Frontiers of Computer Science},
	volume={18},
	number={5},
	pages={185603},
	year={2024},
	publisher={Springer}
}

@inproceedings{chen2023improving,
	title={Improving recommendation fairness via data augmentation},
	author={Chen, Lei and Wu, Le and Zhang, Kun and Hong, Richang and Lian, Defu and Zhang, Zhiqiang and Zhou, Jun and Wang, Meng},
	booktitle={Proceedings of the ACM Web Conference 2023},
	pages={1012--1020},
	year={2023}
}

@inproceedings{pinsage,
	title={Graph convolutional neural networks for web-scale recommender systems},
	author={Ying, Rex and He, Ruining and Chen, Kaifeng and Eksombatchai, Pong and Hamilton, William L and Leskovec, Jure},
	booktitle={Proceedings of the 24th ACM SIGKDD international conference on knowledge discovery \& data mining},
	pages={974--983},
	year={2018}
}

@inproceedings{dgcf,
	title={Disentangled graph collaborative filtering},
	author={Wang, Xiang and Jin, Hongye and Zhang, An and He, Xiangnan and Xu, Tong and Chua, Tat-Seng},
	booktitle={Proceedings of the 43rd international ACM SIGIR conference on research and development in information retrieval},
	pages={1001--1010},
	year={2020}
}

@article{hetrofair,
  author       = {Nemat Gholinejad and
                  Mostafa Haghir Chehreghani},
  title        = {Heterophily-aware fair recommendation using graph convolutional networks},
  journal      = {Neurocomputing},
  volume       = {661},
  pages        = {131956},
  year         = {2026},
  url          = {https://doi.org/10.1016/j.neucom.2025.131956},
  doi          = {10.1016/J.NEUCOM.2025.131956},
  bibsource    = {dblp computer science bibliography, https://dblp.org}
}

@article{wei2023fgcr,
	title={FGCR: Fused graph context-aware recommender system},
	author={Wei, Tianjun and Chow, Tommy WS},
	journal={Knowledge-Based Systems},
	volume={277},
	pages={110806},
	year={2023},
	publisher={Elsevier}
}

@article{riccif2011handbook,
	title={Recommendersystemshandbook},
	author={RicciF, RokachL and ShapiraB, KantorPB},
	journal={Berlin: springerrVerlag},
	year={2011}
}

@article{li2025disentangled,
	title={Disentangled Progressive Negative Sampling for Graph Collaborative Filtering Recommendation},
	author={Li, Hewei and Zhang, Xin and Weng, He and Shen, Yingjie and Cai, Kangkai and Wang, Dongjing and Qin, Zhen and Deng, Shuiguang},
	journal={Knowledge-Based Systems},
	pages={114133},
	year={2025},
	publisher={Elsevier}
}
	
\end{document}